\documentclass[pra,twocolumn, showpacs]{revtex4}
\usepackage{graphicx}
\usepackage{graphics}
\usepackage{amssymb}
\usepackage{epstopdf}
\usepackage{color}
\usepackage{subfigure}
\usepackage{epsfig}
\usepackage{amsmath}
\usepackage{euscript}
\usepackage{float}
\usepackage{color}

\begin{document}
\title{Entanglement and the Interplay between Staggered Fields and Couplings}
\author{Jenny Hide$^{1,2}$, Yoshifumi Nakata$^{2}$ and Mio Murao$^{2,3}$}
\affiliation{The Abdus Salam International Centre for Theoretical Physics
Strada Costiera 11, 34151 Trieste Italy$^1$ \\
Department of Physics, Graduate School of Science, University of Tokyo, Tokyo 113-0033, Japan$^2$\\
Institute for Nano Quantum Information Electronics, University of Tokyo, Tokyo 153-8505, Japan$^3$
}

\pacs{03.65.Ud, 03.67.-a, 75.10.Jm}
\date{\today}

\begin{abstract}
We investigate how the interplay between a staggered magnetic field and staggered
coupling strength affects both ground state and thermal entanglement. Upon
analytically calculating
thermodynamic quantities and the correlation functions for such a system, we
consider both the global Meyer-Wallach measure of entanglement and the concurrence
between pairs of spins. We discover two quantum phase transitions present in the model
and show that the quantum phase transitions are reflected in the behaviour
of the entanglement at zero temperature.
We discover that increasing the alternating field and alternating coupling strength
can actually increase the amount of entanglement present at both zero temperature
and for thermal states of the system. 

\end{abstract}

\maketitle

%%%%%%%%%%%%%%%%%%%%%%%%%%%%%%%%%%%%%%%%%%%%%%%%%%%%%%%%%%%%%%%%%%%%%%%%%%%%%%%%%%%%%%%%
\section{Introduction}
%%%%%%%%%%%%%%%%%%%%%%%%%%%%%%%%%%%%%%%%%%%%%%%%%%%%%%%%%%%%%%%%%%%%%%%%%%%%%%%%%%%%%%%%

Entanglement is an intriguing concept in quantum information, and a resource used
in many quantum computation schemes \cite{manybody}. The Heisenberg coupling in
spin chains has been shown to allow universal quantum computation \cite{Divi}.
Spin chains are also good candidates for quantum wires \cite{Bose}. Thus it is an
important task to explore how the amount of entanglement in such systems changes
under different conditions, to discover whether it can be enhanced and what
causes its destruction.

Many-body entanglement \cite{manybody} is difficult to quantify, so entanglement measures
are often restricted to the pure, zero temperature case, for example von Neumann entropy
\cite{Bennett} or the Meyer-Wallach measure \cite{MW2002}, or to a small number of
possibly mixed qubits such as concurrence \cite{OConnor}. On the other hand, it is,
in general, hard to quantitatively study the entanglement of thermal states
due to the absence of efficient separability criterion for mixed states in many-body systems.
An alternative approach is to use an
entanglement witness which detects rather than measures entanglement. In particular, a
thermodynamic entanglement witness uses thermodynamic quantities derived from
the partition function of the system to detect entanglement \cite{ent_wit,vlat}.

In spin chains, ground state entanglement is often investigated in the context of
quantum phase transitions (QPTs). A QPT is a sudden change in the properties of the
ground state when a parameter of the Hamiltonian such as a magnetic field is varied.
Since it is expected that by investigating QPTs, dramatic changes to physical quantities
at very low temperature would be revealed, QPTs have been intensely studied in spin
systems \cite{Sachdev}. In Ref.~\cite{WSL2004}, it was shown that in general, a singularity
occurs in the ground state entanglement at quantum critical points (QCPs).

In this paper, we investigate the entanglement of a spin system in a
non-uniform magnetic field with non-uniform coupling constants.
Certain solid state systems such as Copper Benzoate have a non-uniform
magnetic field \cite{Oshikawa}, caused by an inhomogeneous Zeeman coupling.
Similarly, examples exist for a non-uniform coupling strength \cite{Abraham,Pincus}.
Some properties of these materials can be captured using the staggered spin chain
we study in this paper.

Entanglement in a staggered magnetic field has been studied previously
using single-site entropy and an entanglement witness \cite{StagHide}.
The effect of such a staggered field on the dynamics
and on the high-fidelity transfer of entanglement has
also been considered \cite{Crooks}; it was found that the staggered field
is almost as efficient as the uniform case.
On the other hand, entanglement in a spin chain with staggered coupling and
magnetic field has been considered only numerically for chains of
finite length \cite{Doronin}.

Here, we show how to calculate two measures of entanglement analytically for such spin chains
with infinite length. We first calculate thermodynamic properties and finite
temperature correlation functions of the spin chain.
Using these results, we show that the system exhibits QPTs at zero temperature
induced by the staggered fields.
We investigate global entanglement of the ground state using the Meyer-Wallach measure,
a measure of multipartite pure states based on bipartite entanglement.
We then investigate both zero and finite temperature entanglement between two
spins using the concurrence.
We find that global entanglement in general increases with an alternating coupling constant.
At zero temperature, the concurrence first increases, reaches a maximum, then
decreases with an alternating
coupling constant. We also find that for certain values of magnetic field, both the
Meyer-Wallach measure and the zero temperature concurrence can be increased by the
alternating magnetic field. At finite temperature, the concurrence can again be increased by certain
values of the staggered fields and couplings.
Further, we examine the entanglement at finite temperature using an entanglement
witness in view of searching for thermal multipartite entanglement.

This paper is organized as follows.
In Section~\ref{Sec:Therm}, we present the Hamiltonian and give its diagonal form, which allows us to
calculate thermodynamical quantities of the system. We compute the correlation functions in Section~\ref{Sec:Corr}.
In Section~\ref{Sec:PropGround}, we investigate the ground state and show that the Hamiltonian exhibits QPTs.
Then, we investigate entanglement at zero temperature and at finite temperature in Section~\ref{Sec:Ent}.
We present concluding remarks in Section~\ref{Sec:Con}.

%%%%%%%%%%%%%%%%%%%%%%%%%%%%%%%%%%%%%%%%%%%%%%%%%%%%%%%%%%%%%%%%%%%%%%%%%%%%%%%%%%
\section{The Hamiltonian and its thermodynamic properties}
%%%%%%%%%%%%%%%%%%%%%%%%%%%%%%%%%%%%%%%%%%%%%%%%%%%%%%%%%%%%%%%%%%%%%%%%%%%%%%%%%%
\label{Sec:Therm}

We consider the thermodynamic limit of the staggered Hamiltonian

\begin{equation}
H = - \sum_{l=1}^N \left[ \frac{J_l}{2} \left( \sigma_l^x \sigma_{l+1}^x + \sigma_l^y
\sigma_{l+1}^y \right) + B_l \sigma_l^z \right], \label{Eq:Hamiltonian}
\end{equation}
where $J_l=J+e^{i \pi l}j$ is the staggered coupling strength and
$B_l=B+e^{i \pi l}b$ is the staggered magnetic field. We refer to $j$ as the
alternating coupling strength and $b$ as the alternating magnetic field.
This spin chain can be diagonalised \cite{Perk} using a Jordan-Wigner transformation,
$a_l = \prod_{m=1}^{l-1} \sigma_m^z \otimes (\sigma_l^x + i \sigma_l^y)/2$,
with anti-commutation relations $\{ a_l,a_m \}=0$ and $\{ a_l^\dagger,a_m \}
= \delta_{l,m}$,  followed
by a Fourier transform, $a_l =N^{-1/2} \sum_k d_k e^{2 i \pi k l /N}$. Next
the Hamiltonian must be rewritten as a sum to $N/2$:

\begin{eqnarray}
H & = & \sum_{k=1}^{N/2} \left( \mu_k^- d_k^\dagger d_k + \mu_k^+
d_{k+N/2}^\dagger d_{k+N/2} \right. \\ \nonumber
& + & \left. \nu_k^+ d_k^\dagger d_{k+N/2} + \nu_k^-
d_{k+N/2}^\dagger d_k - 2 B \textbf{1} \right)
\end{eqnarray}
where $\mu_k^\pm = [2B \pm 2J\cos(2 \pi k/N)]$ and $\nu_k^\pm
= [2b \pm 2ij\sin (2\pi k/N)]$. This can now be diagonalised using a canonical
transformation,

\begin{eqnarray}
d_k & = & \alpha_k \cos \theta_k + \beta_k \sin \theta_k \\ \nonumber
d_{k+N/2} & = & - \alpha_k \sin \theta_k + \beta_k \cos \theta_k
\end{eqnarray}
where $\theta_k$ is determined by $-J \cos(2 \pi k/N) \sin 2\theta_k
+b \cos 2 \theta_k = ij\sin (2 \pi k/N)$. Thus the diagonal form of the
Hamiltonian is

\begin{equation}
H = \sum_{k=1}^{N/2} \left( \lambda_k^+ \alpha_k^\dagger \alpha_k + \lambda_k^-
\beta_k^\dagger \beta_k - 2 B \textbf{1} \right) \label{Eq:Hamiltonian_diagonal}
\end{equation}
where $\lambda_k^\pm = 2B\pm 2\sqrt{J^2\cos^2\left(\frac{2 \pi k}{N}\right)
+b^2+j^2 \sin^2 \left(\frac{2 \pi k}{N}\right)}$. The anti-commutation
relations are now $\{ \alpha_l^\dagger , \alpha_m \} = \{ \beta_l^\dagger ,
\beta_m \} = \delta_{l,m}$ and $\{ \alpha_l , \alpha_m \}=
\{ \beta_l , \beta_m \}=\{ \alpha_l^\dagger , \beta_m \}=
\{ \alpha_l , \beta_m \}= 0$. In all figures in the paper,
we will fix $J$ to $1$.

The partition function, $Z=\mathrm{tr}(e^{-\beta H})$, of this system can be written
$Z = \prod_{k=1}^{N/2} Z_k$ and thus $\ln Z = \sum_{k=1}^{N/2} \ln Z_k$ where
$Z_k = \left[ e^{2 \beta B}+e^{-\beta (\lambda_k^- - 2B)}+e^{-\beta (\lambda_k^+ - 2B)}
+e^{-2\beta B} \right]$ is found from the $k$th term in the Hamiltonian
sum. Explicitly taking the thermodynamic limit, we find

\begin{equation}
\ln Z = \frac{N}{2\pi} \int_0^\pi dq \ln \left[ 4 \cosh\left( \beta \Lambda_q^+ \right)
\cosh\left( \beta \Lambda_q^- \right)\right]
\label{Eq:PartFn1}
\end{equation}
where $\Lambda_q^\pm = B\pm \sqrt{J^2\cos^2 q +b^2+j^2 \sin^2 q}$. From this,
other thermodynamic quantities such as the internal energy, $U=-\frac{\partial}{\partial \beta}
\ln Z$, can be calculated:

\begin{equation}
u:=\frac{U}{N} = - \int_0^\pi \frac{dq}{2\pi} \left[ \Lambda_q^+
\tanh \left( \beta \Lambda_q^+ \right) + \Lambda_q^-
\tanh \left( \beta \Lambda_q^- \right) \right].
\label{Eq:InternalEn}
\end{equation}
The magnetisation, $M=\frac{1}{\beta} \frac{\partial}{\partial B} \ln Z$, is

\begin{equation}
m:=\frac{M}{N} = \int_0^\pi \frac{dq}{2\pi} \left[
\tanh \left( \beta \Lambda_q^+ \right) +
\tanh \left( \beta \Lambda_q^- \right) \right],
\label{Eq:Magnetisation1}
\end{equation}
and the staggered magnetisation , $M_s = \frac{1}{\beta}
\frac{\partial}{\partial b} \ln Z$, is

\begin{equation}
m_s:=\frac{M_s}{N} = \int_0^\pi \frac{dq}{2\pi}
\frac{b \left[ \tanh \left( \beta \Lambda_q^+ \right) -
\tanh \left( \beta \Lambda_q^- \right)
\right]}{\sqrt{J^2\cos^2 q +b^2+j^2 \sin^2 q}}.
\label{Eq:StaggeredMag}
\end{equation}

%%%%%%%%%%%%%%%%%%%%%%%%%%%%%%%%%%%%%%%%%%%%%%%%%%%%%%%%%%%%%%%%%%%%%%%%%%%%%%%%%%%%%%%
\section{Correlation functions}
%%%%%%%%%%%%%%%%%%%%%%%%%%%%%%%%%%%%%%%%%%%%%%%%%%%%%%%%%%%%%%%%%%%%%%%%%%%%%%%%%%%%%%%
\label{Sec:Corr}

We will use the correlation functions of the system to calculate each
measure of entanglement. Since $[H,\sum_l \sigma_l^z ]=0$, the only non-zero
correlation functions of this spin chain are $\langle \sigma_l^x \sigma_{l+R}^x + \sigma_l^y
\sigma_{l+R}^y \rangle$, $\langle \sigma_l^z \sigma_{l+R}^z \rangle$
and $\langle \sigma_l^z \rangle$.
The Hamiltonian is semi-translationally invariant in that all odd and all even sites
can be considered identical in the thermodynamic limit.
Due to this semi-translational invariance, when $R$ is even,
$\langle \sigma_l^z \rangle = \langle \sigma_{l+R}^z\rangle$.

We calculate each of the correlation functions following the
method in \cite{Barouch}. We find these for any $R$. Generally,
$\langle \sigma_l^x \sigma_{l+R}^x + \sigma_l^y \sigma_{l+R}^y \rangle
= 2\langle a_l^\dagger \prod_{m=l+1}^{R-1} (\textbf{1}-2 a_m^\dagger  a_m)
a_{l+R} + a_{l+R}^\dagger  \prod_{m=l+1}^{R-1} (\textbf{1}-2 a_m^\dagger
a_m) a_l\rangle$,
$\langle \sigma_l^z \sigma_{l+R}^z \rangle =\langle (\textbf{1}-2 a_l^\dagger  a_l)
(\textbf{1}-2 a_{l+R}^\dagger  a_{l+R})\rangle $ and $\langle \sigma_l^z \rangle
=  \langle \textbf{1}-2 a_l^\dagger  a_l\rangle$.

The auto-correlation functions of this model have been found analytically
in the infinite temperature limit \cite{PerkInfT}, and the time dependent $\langle \sigma_l^z (t)
\sigma_{l+R}^z\rangle$ has been found, also analytically, for arbitrary temperature \cite{PerktCorr}.
However, the general equilibrium correlation functions have not been calculated
previously.

Since the spin chain is a free fermion model and semi-translationally invariant,
Wick's theorem can be applied to the total correlation functions.
Thus we can rewrite each of the above equations in terms of two point
correlation functions. For example, the $zz$ correlation function is
$\langle (\textbf{1}-2 a_l^\dagger  a_l)
(\textbf{1}-2 a_{l+R}^\dagger  a_{l+R})\rangle =  \langle \sigma_l^z \rangle
\langle \sigma_{l+R}^z \rangle- G_{l,R}^2$ where
we have defined $G_{l,R}=  -\langle a_l^\dagger a_{l+R} - a_l a_{l+R}^\dagger
\rangle$.
We use the notation $G_{l,R} = G_{R}^0 +e^{i\pi l}G_{R}^s$ and $\langle \sigma_l^z \rangle
= \langle \sigma^z \rangle^0 +e^{i\pi l} \langle \sigma^z \rangle^s$
where $G_{R}^0 = - \frac{1}{N} \sum_l \langle a_l^\dagger a_{l+R} - a_l a_{l+R}^\dagger
\rangle$ and $G_{R}^s = -\frac{1}{N} \sum_l e^{i \pi l} \langle a_l^\dagger a_{l+R} -
a_l a_{l+R}^\dagger \rangle$ (similarly for $\langle \sigma_l^z \rangle$).
Note that we have treated $\langle \sigma_l^z \rangle$
separately to $G_{l,R}$ despite the fact that $R=0$ should give us $\langle \sigma_l^z \rangle$. The
reason for this will become clear below.

Using the Jordan-Wigner transformation and Fourier transform, and then summing
to $N/2$ (the additional canonical transformation is unnecessary), we find we can write

\begin{eqnarray}
\langle \sigma_l^z \rangle^0 & = & \frac{1}{N}
\sum_{k=1}^{N/2} \left[ \langle \textbf{1} - 2 d_k^\dagger d_k \rangle +
\langle \textbf{1} - 2 d_{k+\frac{N}{2}}^\dagger d_{k+\frac{N}{2}} \rangle \right] \nonumber \\
\langle \sigma_l^z \rangle^s & = & -\frac{2}{N}
\sum_{k=1}^{N/2} \left[ \langle d_k^\dagger d_{k+\frac{N}{2}} \rangle +
\langle d_{k+\frac{N}{2}}^\dagger d_{k} \rangle \right],
\label{Eq:AfterTrans1}
\end{eqnarray}
for $\langle \sigma_l^z \rangle$ and

\begin{eqnarray}
G_{R}^0 & = & \frac{1}{N}
\sum_{k=1}^{N/2} \cos \left(\frac{2 \pi k R}{N}\right) \left[ \right. \langle \textbf{1} -
2 d_k^\dagger d_k \rangle \nonumber \\
& + & e^{i\pi R} \langle \textbf{1} - 2 d_{k+\frac{N}{2}}^\dagger
d_{k+\frac{N}{2}} \rangle \left. \right]
\end{eqnarray}

\begin{equation}
G_{R}^s =  \frac{2i}{N} \sum_{k=1}^{N/2} \sin \left(\frac{2 \pi k R}{N}\right)
\left[ \langle d_k^\dagger d_{k+\frac{N}{2}} \rangle +
e^{i\pi R} \langle d_{k+\frac{N}{2}}^\dagger d_{k} \rangle \right], \nonumber
\label{Eq:AfterTrans2}
\end{equation}
for $G_{l,R}$.
We know the thermodynamic forms of $\langle \sigma_l^z \rangle^0 = m$ (Eq.~\ref{Eq:Magnetisation1}),
$\langle \sigma_l^z \rangle^s = m_s$ (Eq.~\ref{Eq:StaggeredMag}), and also of
$G_{l,1}=-\langle \sigma_l^x \sigma_{l+1}^x + \sigma_l^y \sigma_{l+1}^y \rangle/2$
which can be calculated by differentiating the partition
function with respect to the coupling strengths, $G_{l,1}^0= -\frac{1}{N\beta} \frac{\partial}{\partial J}
\ln Z$ and $G_{l,1}^s= -\frac{1}{N\beta} \frac{\partial}{\partial j} \ln Z$. Thus we have

\begin{eqnarray}
G_{1}^0 & = & - \int_0^\pi \frac{dq}{2\pi} \frac{J \cos^2 q  \left[ \tanh (\beta
\Lambda^+) -  \tanh (\beta \Lambda^-) \right]}{\sqrt{J^2 \cos^2 q +b^2 +j^2 \sin^2 q}} \nonumber \\
G_{1}^s & = & -\int_0^\pi \frac{dq}{2\pi} \frac{j \sin^2 q  \left[ \tanh (\beta
\Lambda^+) -  \tanh (\beta \Lambda^-) \right]}{\sqrt{J^2 \cos^2 q +b^2 +j^2 \sin^2 q}}, \nonumber
\end{eqnarray}
The thermodynamic expressions for $G_{l,R}$ are found directly from the above equations.
For even $R$, we compare the correlation function form of $G_{l,R}$ to that of $\langle
\sigma_l^z \rangle$, noticing that they are similar. Since we know the thermodynamic form
of $\langle \sigma_l^z \rangle$, we can also determine the thermodynamic form of $G_{l,R}$:

\begin{eqnarray}
G_{R}^0 & = & \int_0^\pi \frac{dq}{2\pi} \cos(qR) \left[ \tanh (\beta \Lambda^+)
+  \tanh (\beta \Lambda^-) \right] \\
G_{R}^s  & = & -i \int_0^\pi \frac{dq}{2\pi} \frac{b \sin(qR)  \left[ \tanh (\beta
\Lambda^+) -  \tanh (\beta \Lambda^-) \right]}{\sqrt{J^2 \cos^2 q +b^2 +j^2 \sin^2 q}}. \nonumber
\label{Eq:GREven}
\end{eqnarray}
We note that for $R$ even, $G_R^s = 0$. For odd $R$, we instead compare the correlation function
form of $G_{l,R}$ to $G_{l,1}$, again noticing they are similar. Using the thermodynamic form
of $G_{l,1}$, we determine the thermodynamic form of $G_{l,R}$:

\begin{eqnarray}
G_{R}^0 & = & -\int_0^\pi \frac{dq}{2\pi} \cos(qR) \frac{J \cos q  \left[ \tanh (\beta
\Lambda^+) -  \tanh (\beta \Lambda^-) \right]}{\sqrt{J^2 \cos^2 q +b^2 +j^2 \sin^2 q}} \nonumber \\
G_{R}^s & = & -\int_0^\pi \frac{dq}{2\pi} \sin (qR) \frac{j \sin q  \left[ \tanh (\beta
\Lambda^+) -  \tanh (\beta \Lambda^-) \right]}{\sqrt{J^2 \cos^2 q +b^2 +j^2 \sin^2 q}}, \nonumber
\label{Eq:GROdd}
\end{eqnarray}
The differences between odd and even $R$ are due to the presence of the $e^{i \pi R}$ term
in the correlation function forms of $G_{l,R}$.

%%%%%%%%%%%%%%%%%%%%%%%%%%%%%%%%%%%%%%%%%%%%%%%%%%%%%%%%%%%%%%%%%%%%%%%%%%%%%%%%%%%%%%%
\section{Properties of the ground state}
%%%%%%%%%%%%%%%%%%%%%%%%%%%%%%%%%%%%%%%%%%%%%%%%%%%%%%%%%%%%%%%%%%%%%%%%%%%%%%%%%%%%%%%
\label{Sec:PropGround}

\begin{figure}[tb]
\centering
\begin{tabular}{cc}
 \includegraphics[width=13em, clip]{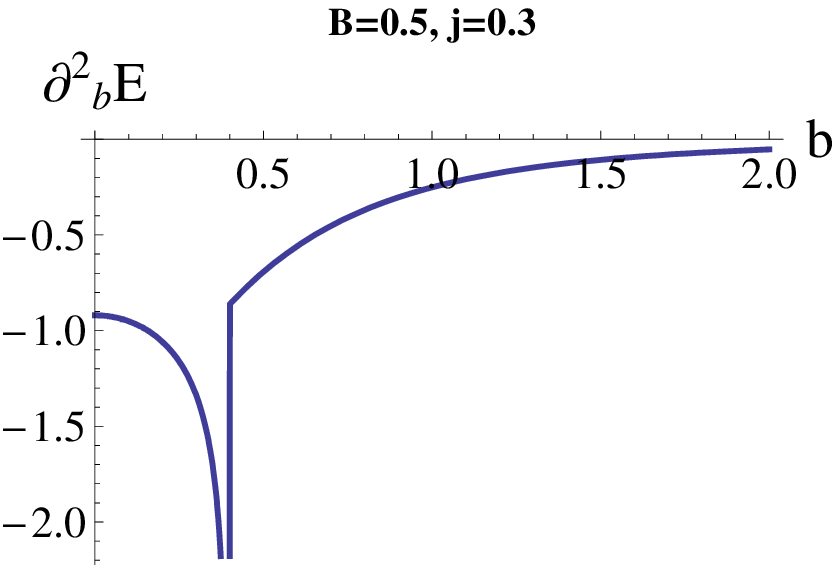} &
 \includegraphics[width=13em, clip]{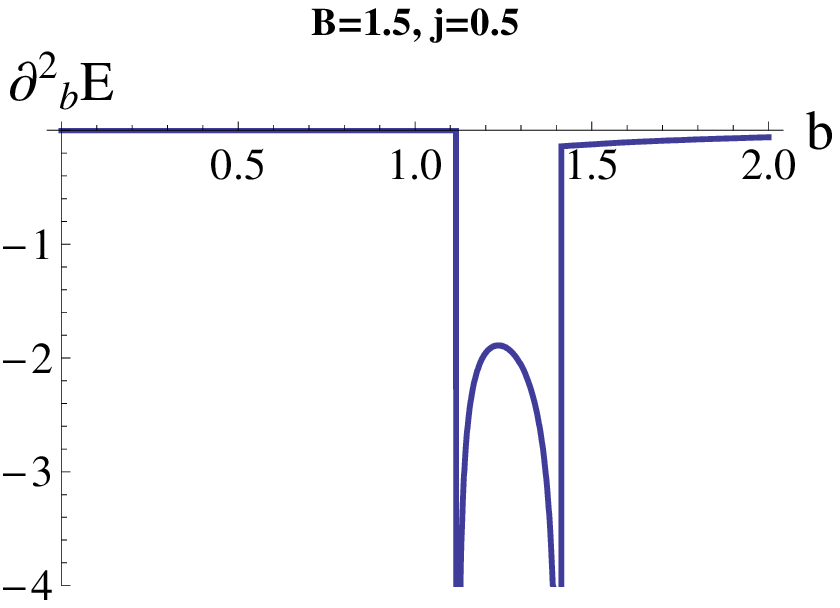}\\
  \footnotesize \ \ \ \ (a) &  \footnotesize \ \ \ \ (b)\\
  \end{tabular}
  \caption{Second derivative of the ground energy with respect to $b$. The
  figures (a) and (b) are for $(B,j)=(0.5,0.3)$ and $(B,j)=(1.5,0.5)$,
   respectively. It is observed that the second derivative diverges at the QCPs
   $B=\sqrt{J^2 + b^2}$ and $B=\sqrt{j^2+b^2}$.}
	\label{Fig:DerivativeOfGroundEnergy}
\end{figure}

\begin{figure}[tb]
\centering
\begin{tabular}{cc}
 \includegraphics[width=14em, clip]{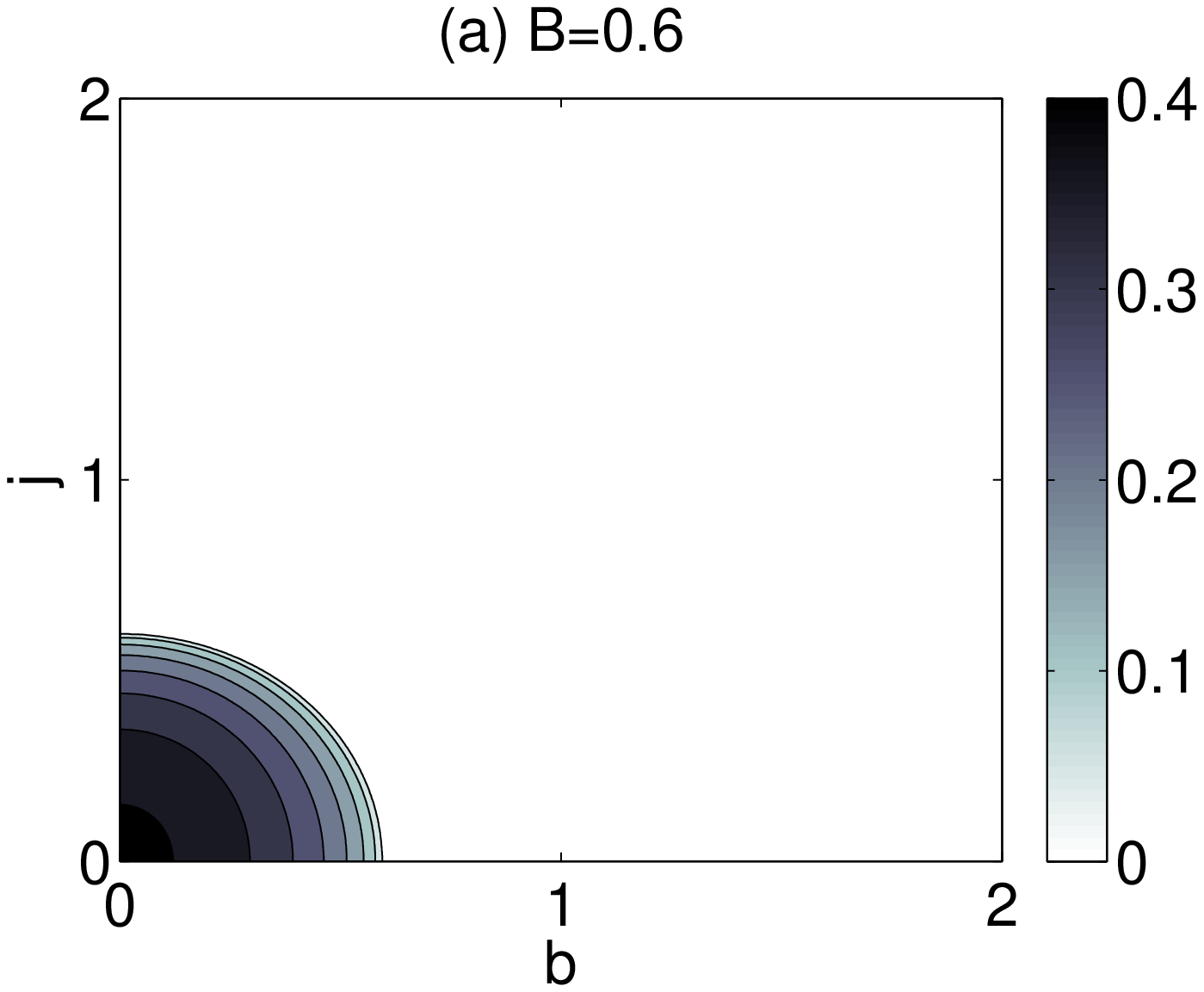} &
 \includegraphics[width=14em, clip]{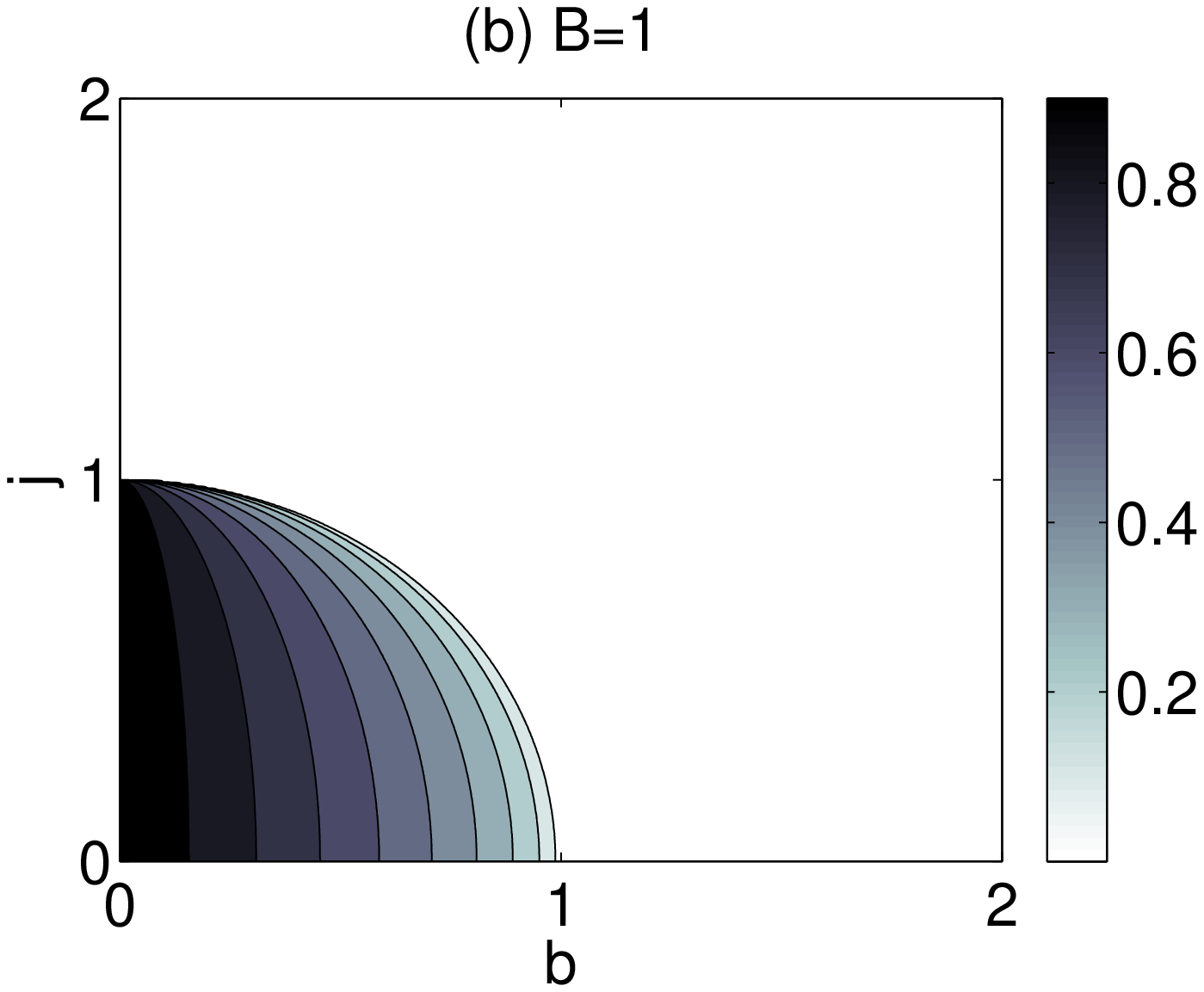}\\
  \includegraphics[width=14em, clip]{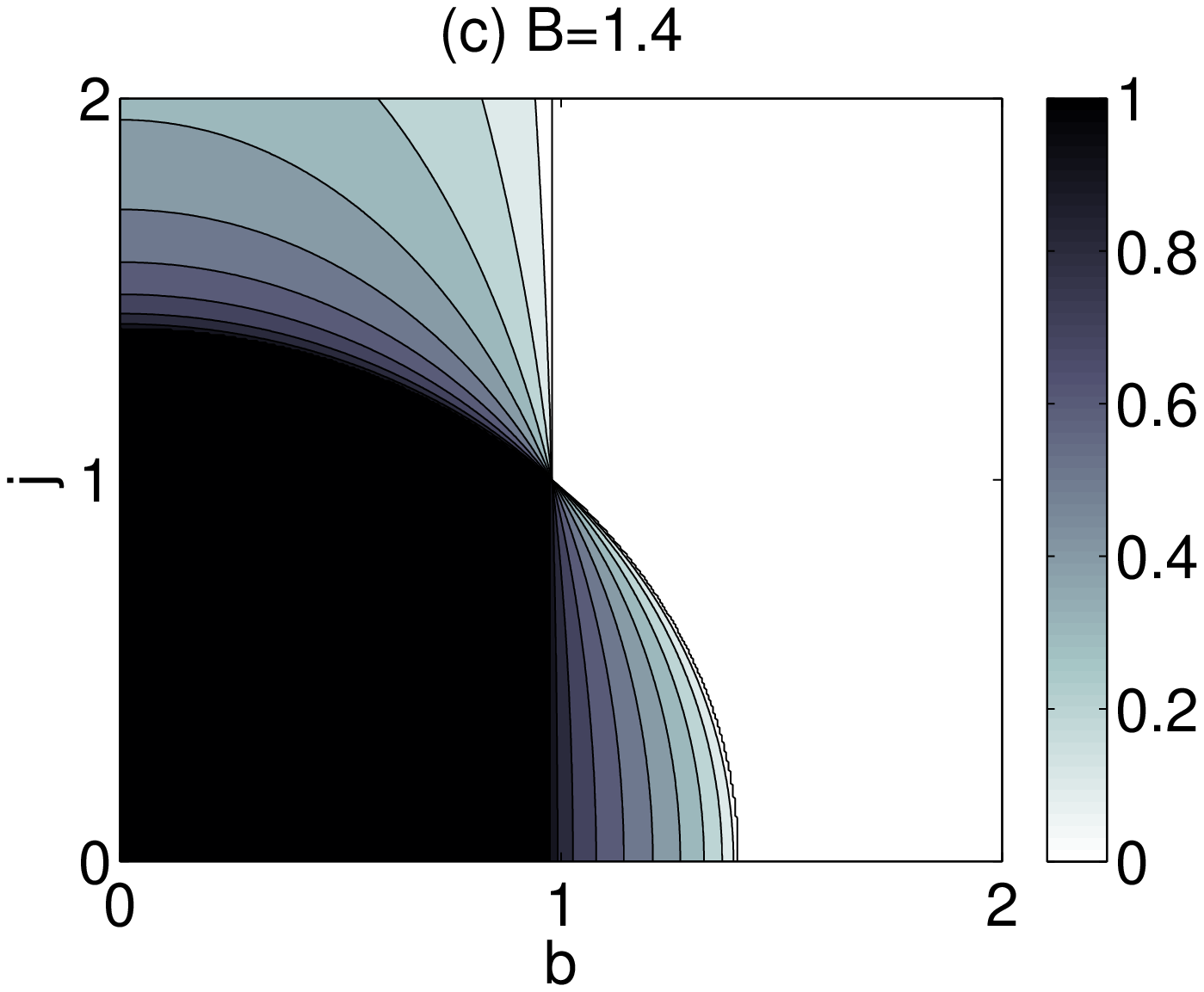}&
  \end{tabular}
  \caption{Magnetization at zero temperature. The figures (a), (b) and (c)
  show $B=0.6$, $B=1$ and $B=1.4$, respectively. The magnetization changes
  non-smoothly at the quantum critical points $B=\sqrt{J^2 + b^2}$ and
  $B=\sqrt{j^2+b^2}$. }
	\label{Fig:Magnetization}
\end{figure}

In this section, we investigate the properties of the ground state. Without the
alternating coupling strength and the alternating magnetic field, the Hamiltonian is
referred to as an XX model with a transverse magnetic field.
It is well known that the XX model has a second order QPT at $B=J$ \cite{Takahashi}.
We investigate the ground state energy as
a function of the alternating coupling strength and the alternating magnetic field,
and see the QPTs induced by them.

We first define $Q$ as
\begin{equation}
Q=\{ Q \in [0, \pi] | \Lambda^-_q < 0 \}.
\end{equation}
For simplicity, we also introduce two functions,
\begin{align}
\Theta(q) &= \sqrt{J^2 \cos^2 q + b^2 + j^2 \sin^2 q},\\
\Xi &= \arccos \sqrt{\frac{B^2 - b^2 - j^2}{J^2-j^2}}.
\end{align}
By using the internal energy per site $u$ given by Eq.~\eqref{Eq:InternalEn},
the ground state energy per site is obtained by taking the limit such as
\begin{align}
\epsilon_g &= - \lim_{\beta \rightarrow \infty} \int_0^{\pi} \frac{dq}{2\pi}[\Lambda^+_q
\tanh ( \beta \Lambda^+_q) + \Lambda^-_q \tanh ( \beta \Lambda^-_q)]  \notag \\
&= -\int_{q \notin Q} \frac{dq}{2\pi} [\Lambda^+_q + \Lambda^-_q] -\int_{q \in Q}
\frac{dq}{2\pi} [\Lambda^+_q - \Lambda^-_q] \notag \\
&=- B+ \frac{1}{\pi} \int_{q \in Q} [ B -  \Theta(q)] dq. \label{Eq:GroundEnergy}
\end{align}
In Appendix~\ref{App:ZeroTemp}, the region $Q$ is analytically obtained and exact
expressions of the ground state energy are shown.
In order to visualize the QPTs,
the second derivative of the ground state in terms of $b$ for $(B,j)=(0.5,0.3)$ and
$(B,j)=(1.5,0.5)$ is given in Fig.~\ref{Fig:DerivativeOfGroundEnergy}.
It is observed that the derivative diverges at the points $B=\sqrt{J^2 + b^2}$ and $B=\sqrt{j^2+b^2}$,
both of which lead to second order QPTs.

The QPTs are observed more clearly by looking at the magnetic susceptibility at zero temperature.
In Appendix~\ref{App:ZeroTemp}, the magnetization at zero temperature is explicitly calculated, which is
a function of $\Xi$. Since the derivative of $\Xi$ by $B$ and $b$ includes a factor $1/\sqrt{B^2-b^2-j^2}\sqrt{-B^2+J^2+b^2}$,
the derivative diverges at the points $B=\sqrt{J^2 + b^2}$ and $B=\sqrt{j^2+b^2}$, which implies QPTs.
To demonstrate this, we also plot the magnetization at zero temperature in Fig.~\ref{Fig:Magnetization}.
We can clearly see that the magnetization of the ground state changes non-smoothly at both QCPs.

We note that, when $j=b=0$ corresponding to the XX model with a transverse magnetic field,
$B=\sqrt{J^2 + b^2}$ reduces to the QCP of the XX model, $B=J$.
On the other hand, QPTs at $B=\sqrt{j^2+b^2}$ do not appear in the XX model.
Hence this QCP can be regarded as being induced by the staggered
nature of the spin chain.

%%%%%%%%%%%%%%%%%%%%%%%%%%%%%%%%%%%%%%%%%%%%%%%%%%%%%%%%%%%%%%%%%%%%%%%%%%%%%%%%%%%%%%%
\section{Entanglement}
%%%%%%%%%%%%%%%%%%%%%%%%%%%%%%%%%%%%%%%%%%%%%%%%%%%%%%%%%%%%%%%%%%%%%%%%%%%%%%%%%%%%%%%
\label{Sec:Ent}

We study the entanglement properties of the spin chain in three different ways. First
we define the Meyer-Wallach measure and the concurrence and discuss how to calculate
each for the staggered Hamiltonian. Next we consider how both entanglement measures
behave at zero temperature, and then how a finite temperature affects the concurrence.
Finally we consider a thermodynamic entanglement witness in
an attempt to detect thermal entanglement which the measures miss such as multipartite
entanglement.

\subsection{Meyer-Wallach measure and Concurrence}

First, we show how to calculate the Meyer-Wallach measure and the concurrence from the thermodynamic
quantities calculated in Section~\ref{Sec:Therm} and correlation functions
in Section~\ref{Sec:Corr}.

For a pure state $|\Phi \rangle$ of an $N$-spin system, the Meyer-Wallach measure is defined by
\begin{equation}
E_{\mathrm{MW}}(|\Phi \rangle)=2-\frac{2}{N} \sum_{i=1}^N {\rm Tr} \rho_i^2, \notag
\end{equation}
where
$\rho_i:={\rm Tr}_{\neg i} |\Phi\rangle \langle \Phi|$ is the reduced density matrix at the
$i$-th spin~ \cite{MW2002, B2003}. (The partial trace ${\rm Tr}_{\neg i}$ is taken over all
degrees of freedom except the $i$-th spin.)
The Meyer-Wallach measure takes values between $0$ and $1$. The minimum is achieved if and only if the state
is separable while the maximum is given by states which are local unitary equivalent to the GHZ state.

We calculate the entanglement of the ground state found by the Meyer-Wallach measure.
Since the Hamiltonian preserves the magnetization, i.e., $[H,\sum_l \sigma_l^z ]=0$,
the reduced density matrix of a spin at site $l$, $\rho_l$, has only diagonal elements
such that $\rho_l={\rm diag}\{\frac{1+\langle \sigma_l^z \rangle_g}{2},
\frac{1-\langle \sigma_l^z \rangle_g}{2} \}$ where $\langle \sigma_l^z \rangle_g := \langle g
| \sigma_l^z | g \rangle$ is the expectation value of $\sigma_l^z$ by the ground state $|g \rangle$.
By substituting this and using the fact that the Hamiltonian is semi-translationally invariant, the Meyer-Wallach measure of the ground state is obtained as
\begin{equation}
E_{\mathrm{MW}}(| g \rangle) = 1- \frac{1}{2}( \langle \sigma_{even}^z \rangle_g^2
+\langle \sigma_{odd}^z \rangle_g^2 ). \label{Eq:EMWGround}
\end{equation}
where $\langle \sigma_{even}^z \rangle_g = \langle g | \sigma_{2l}^z | g \rangle$ and
$\langle \sigma_{odd}^z \rangle_g = \langle g | \sigma_{2l+1}^z | g \rangle$ for any $l$.
The $\langle \sigma_{even}^z \rangle$ and $\langle \sigma_{odd}^z \rangle$ are obtained from
the magnetization $m$ given by Eq.~\eqref{Eq:Magnetisation1} and $m_s$ given by
Eq.~\eqref{Eq:StaggeredMag} such as
\begin{align}
\langle \sigma_{even}^z \rangle_g &= \lim_{\beta \rightarrow \infty} [m + m_s], \\
\langle \sigma_{odd}^z \rangle_g &=\lim_{\beta \rightarrow \infty} [m-m_s].
\end{align}
The exact expressions of the ground state energy, the magnetization at zero temperature
and the Meyer-Wallach measure are given in Appendix~\ref{App:ZeroTemp}.
Note that the Meyer-Wallach measure is a measure of entanglement only for pure states,
and thus is meaningful for investigation of entanglement in the
ground state but not that of thermal states.

The concurrence, ${\mathcal C}$, between two spins \cite{OConnor} is an entanglement measure
for both pure and mixed states, and can therefore be used at finite temperatures.
It is given by

\begin{equation}
{\mathcal C}(\rho) = \max \{0,\lambda_1 - \lambda_2 - \lambda_3 - \lambda_4 \},
\end{equation}
where the $\lambda_i$s are the square roots of the eigenvalues
of the matrix $\rho \tilde{\rho}$ with $\tilde{\rho} = (\sigma^y
\otimes \sigma^y) \rho^* (\sigma^y \otimes
\sigma^y)$, and satisfy $\lambda_1 \geq \lambda_2 \geq \lambda_3 \geq \lambda_4$.
Again using $[H,\sum_l \sigma_l^z ]=0$,
the concurrence between two spins at sites $l$ and $l+R$ is
${\mathcal C} (\rho_{l,l+R})=2\max\{0,|z|-\sqrt{vy}\}$  where
$z=\frac{1}{4}\langle \sigma_l^x \sigma_{l+R}^x + \sigma_l^y
\sigma_{l+R}^y \rangle$, and $vy = \frac{1}{16}[ (1+
\langle \sigma_l^z \sigma_{l+R}^z \rangle)^2 -(\langle \sigma_l^z
\rangle+ \langle \sigma_{l+R}^z\rangle)^2]$.
Due to the semi-translational invariance of the Hamiltonian, the concurrence is
the same for all odd sites and for all even sites. The concurrence is 
zero if and only if the state of the two spins is separable, and is one
when they are maximally entangled.
Although we could calculate the concurrence for any $R$, we concentrate on the
nearest neighbour, ${\mathcal C}_1$ with $R=1$, and the next nearest neighbour,
${\mathcal C}_2$ with $R=2$, concurrence
since for large $R$ the concurrence is infinitesimal. Using
$\langle \sigma_l^z \sigma_{l+R}^z \rangle = \langle \sigma_l^z \rangle
\langle \sigma_{l+R}^z \rangle - G_{l,R}^2$, the nearest neighbour concurrence is

\begin{equation}
{\mathcal C}_1 = \max \left\{ 0, \left| G_{l,1} \right| - \frac{1}{2}
\sqrt{\left(1+ \langle \sigma_l^z \sigma_{l+1}^z \rangle \right)^2 -\left(
2 \langle \sigma^z \rangle^0  \right)^2} \right\},
\end{equation}
and the next nearest neighbour concurrence is

\begin{eqnarray}
{\mathcal C}_2 & = & \max \bigg\{ 0, \left| G_{l,1} G_{l+1,1}-G_{l,2}\langle
\sigma_{l+1}^z \rangle \right|  \bigg. \\ \nonumber
& - & \left. \frac{1}{2}
\sqrt{\left(1+ \langle \sigma_l^z \sigma_{l+2}^z \rangle \right)^2 -\left(
2 \langle \sigma_l^z \rangle  \right)^2} \right\},
\end{eqnarray}
remembering that $G_R$ is different for odd and for even values of $R$.

When $J=j$, the total coupling strength between nearest neighbour sites for odd $l$
is zero, while for even $l$, it is $2J$. Thus, at any temperature, there is no
entanglement between nearest neighbours for odd sites, and at zero temperature (and
magnetic fields), the chain consists of $N/2$ maximally entangled singlet states.
This is an example of dimerisation. As a consequence of this, there is also no
concurrence at $J=j$ for both odd and even sites for any $R>1$.

We note that unlike the Meyer-Wallach measure, the concurrence
is not directly related to the total amount of entanglement contained in a pure state.
Hence, there is no guarantee that the Meyer-Wallach measure and the concurrence will
behave similarly. For instance, for the GHZ state, the Meyer-Wallach measure is one
but the concurrence between any two spins is zero.

%%%%%%%%%%%%%%%%%%%%%%%%%%%%%%%%%%%%%%%%%%%%%%%%%%%%%%%%%%%%%%%%%%%%%%%%%%%%%%%%%%%%%%%
\subsection{Zero temperature}
%%%%%%%%%%%%%%%%%%%%%%%%%%%%%%%%%%%%%%%%%%%%%%%%%%%%%%%%%%%%%%%%%%%%%%%%%%%%%%%%%%%%%%%
\label{SubSec:Meyer}

\begin{figure}[t]
\centering
\begin{tabular}{cc}
\epsfig{file=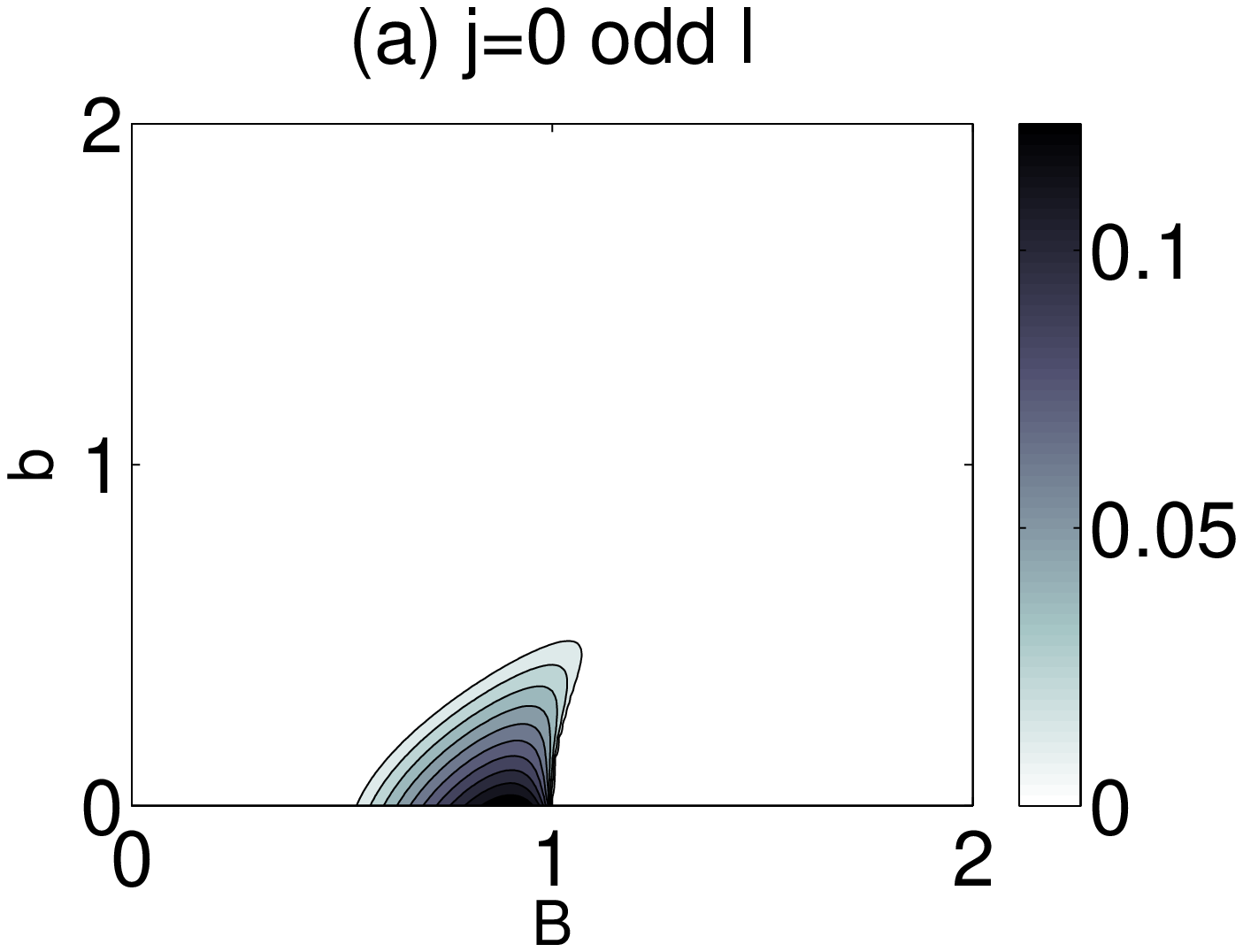,width=0.5\columnwidth,clip=} &
\epsfig{file=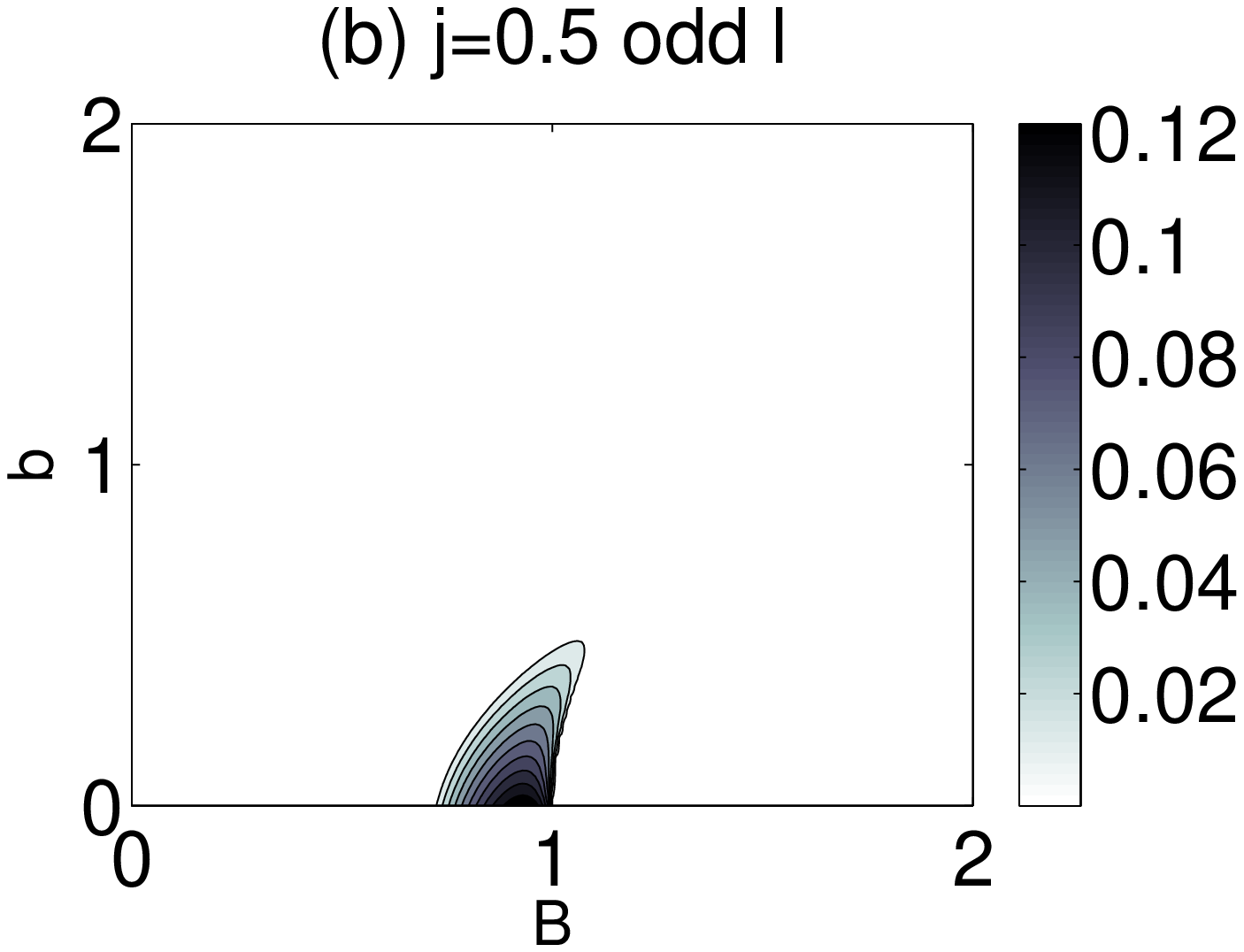,width=0.5\columnwidth,clip=} \\
\epsfig{file=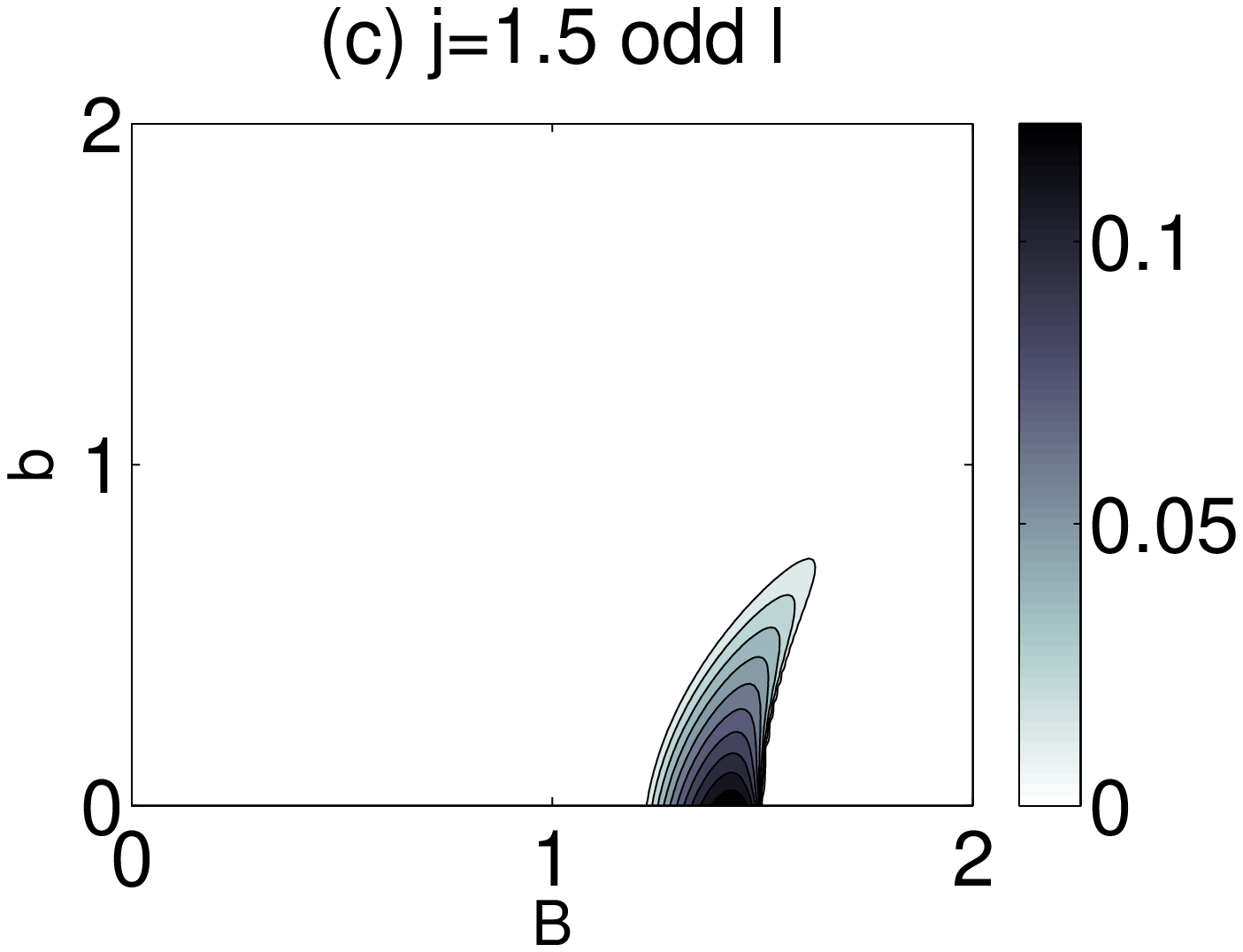,width=0.5\columnwidth,clip=} &
\epsfig{file=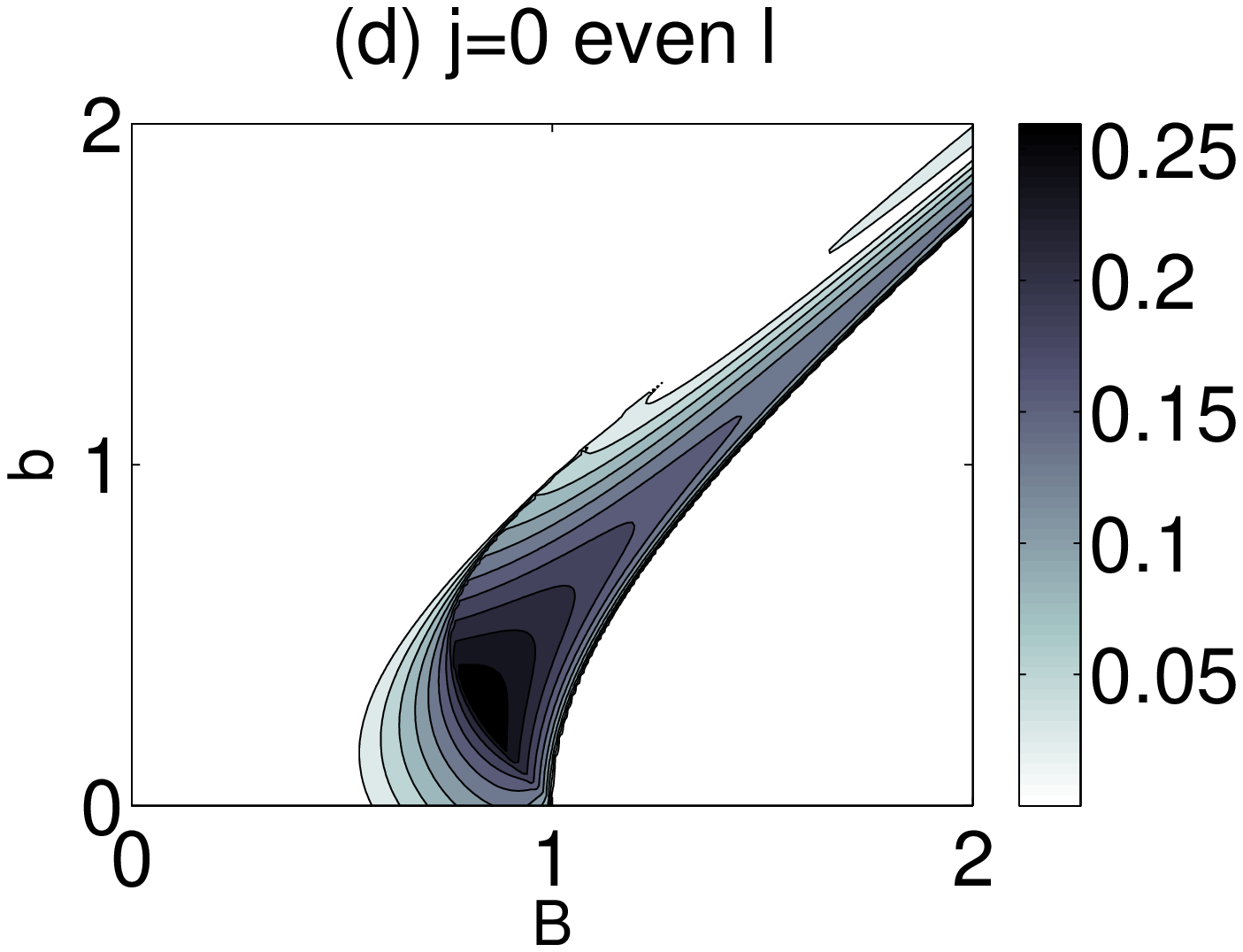,width=0.5\columnwidth,clip=} \\
\epsfig{file=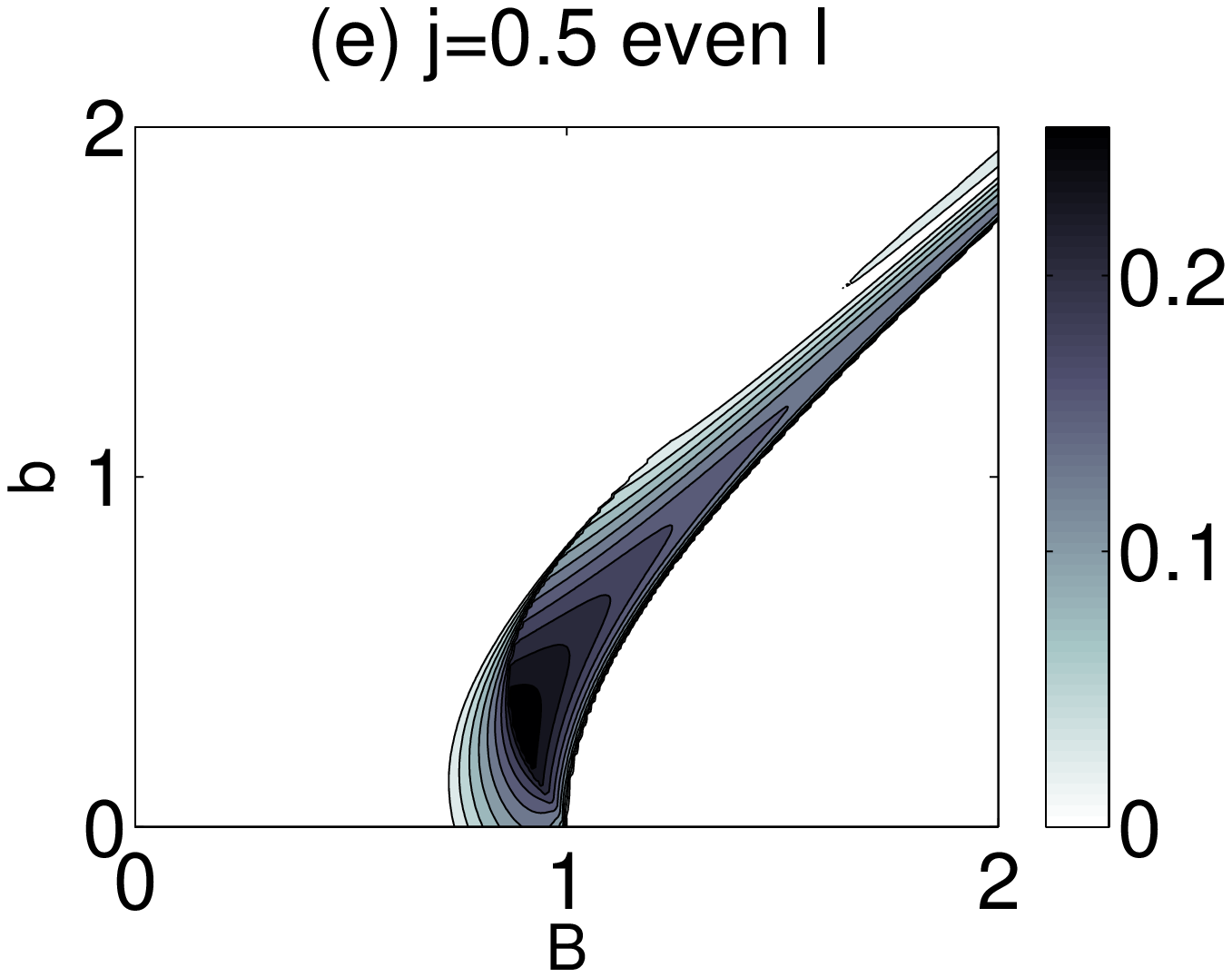,width=0.5\columnwidth,clip=} &
\epsfig{file=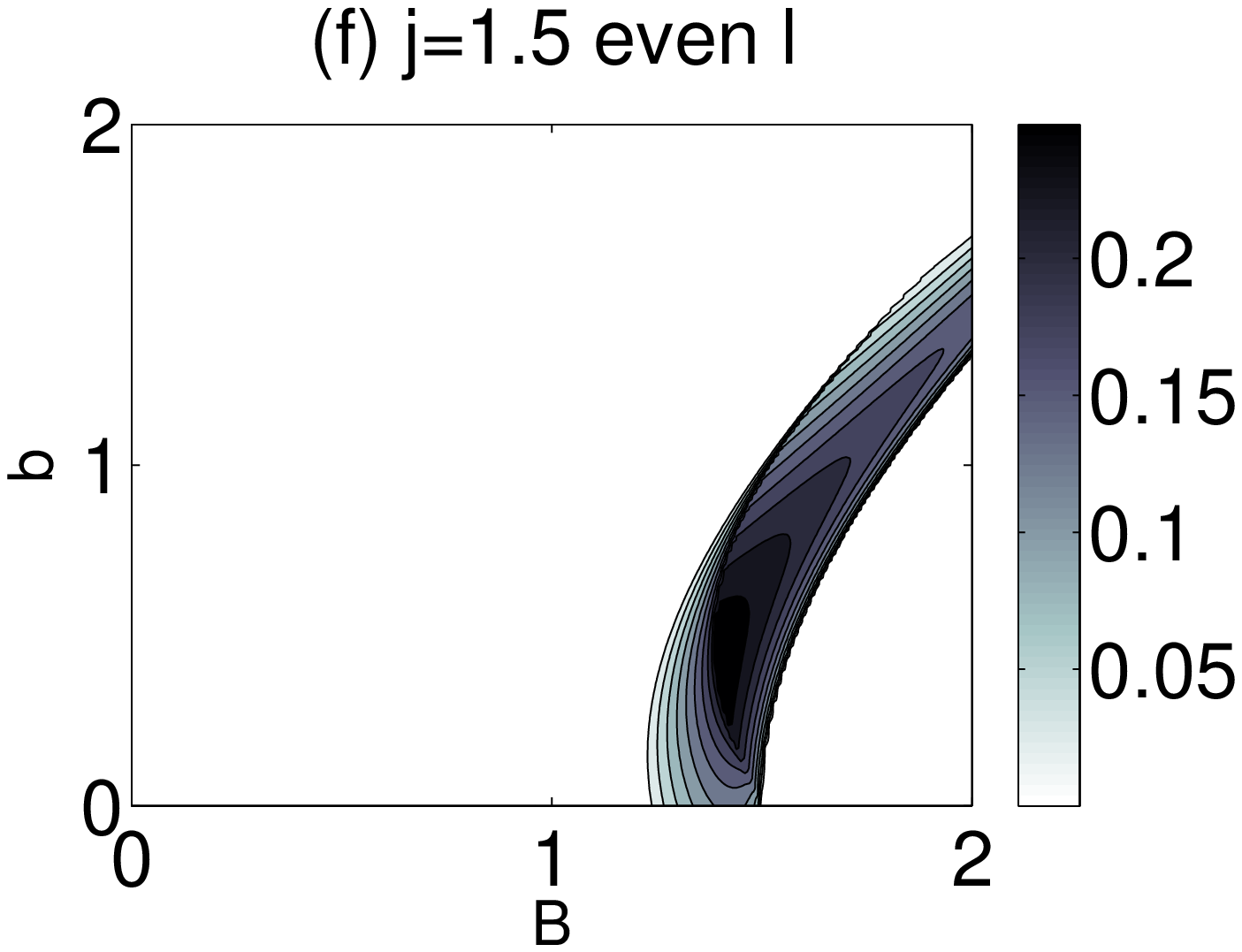,width=0.5\columnwidth,clip=}
\end{tabular}
\caption{Next nearest neighbour concurrence as $T\rightarrow 0$.
Concurrence is zero when $j=1$ as discussed in the text.}
\label{Fig:NNNConcT0j}
\end{figure}

In this section, we study entanglement at zero temperature using the Meyer-Wallach measure and the
concurrence as defined above.
Figs.~\ref{Fig:MeyerWallach2} and~\ref{Fig:MeyerWallach3} show the Meyer-Wallach measure,
Figs.~\ref{Fig:NNConcT0j} and~\ref{Fig:NNConcT0b} the nearest neighbour (NN) concurrence
for both odd and even sites, and Fig.~\ref{Fig:NNNConcT0j} the next nearest neighbour (NNN)
concurrence.

\subsubsection{Quantum phase transitions}

We first discuss the Meyer-Wallach measure of the ground state.
See Appendix~\ref{App:ZeroTemp} for the detailed calculation.
In Figs.~\ref{Fig:MeyerWallach2} and~\ref{Fig:MeyerWallach3}, it is observed that
the Meyer-Wallach measure changes non-smoothly at the QCPs.

Conversely, considering the plots for concurrence at zero temperature, the
quantum phase transitions present in the staggered model are not always evident.
In particular, when plotting $B$ against $b$ for odd sites in Fig.~\ref{Fig:NNConcT0j},
only the curve $B=\sqrt{J^2+b^2}$ can be seen in (b), while only $B=\sqrt{j^2+b^2}$ can be
seen in (c). However, changes in the concurrence can be observed at both QCPs for
even $l$ in the same figure, and for both odd and even sites in Fig.~\ref{Fig:NNConcT0b} where
we plot $B$ against $j$.

These results are consistent with the results in Ref.~\cite{WSL2004}, where it is shown that
entanglement of the ground state behaves singularly around QCPs in general.

\subsubsection{Effects of $j$ and $b$ on entanglement}

\begin{figure}[tb]
\centering
\begin{tabular}{cc}
 \includegraphics[width=13em, clip]{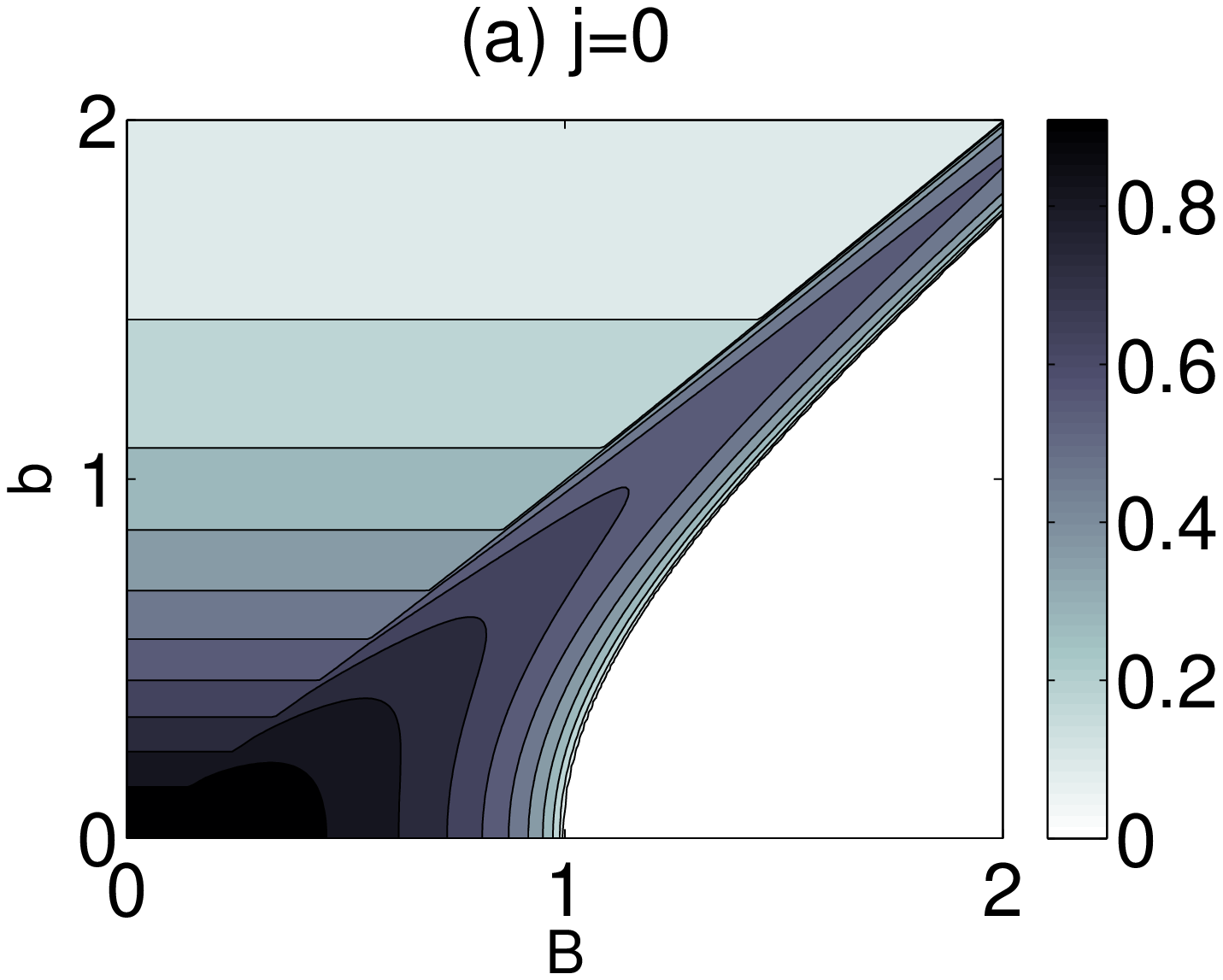} &
 \includegraphics[width=13em, clip]{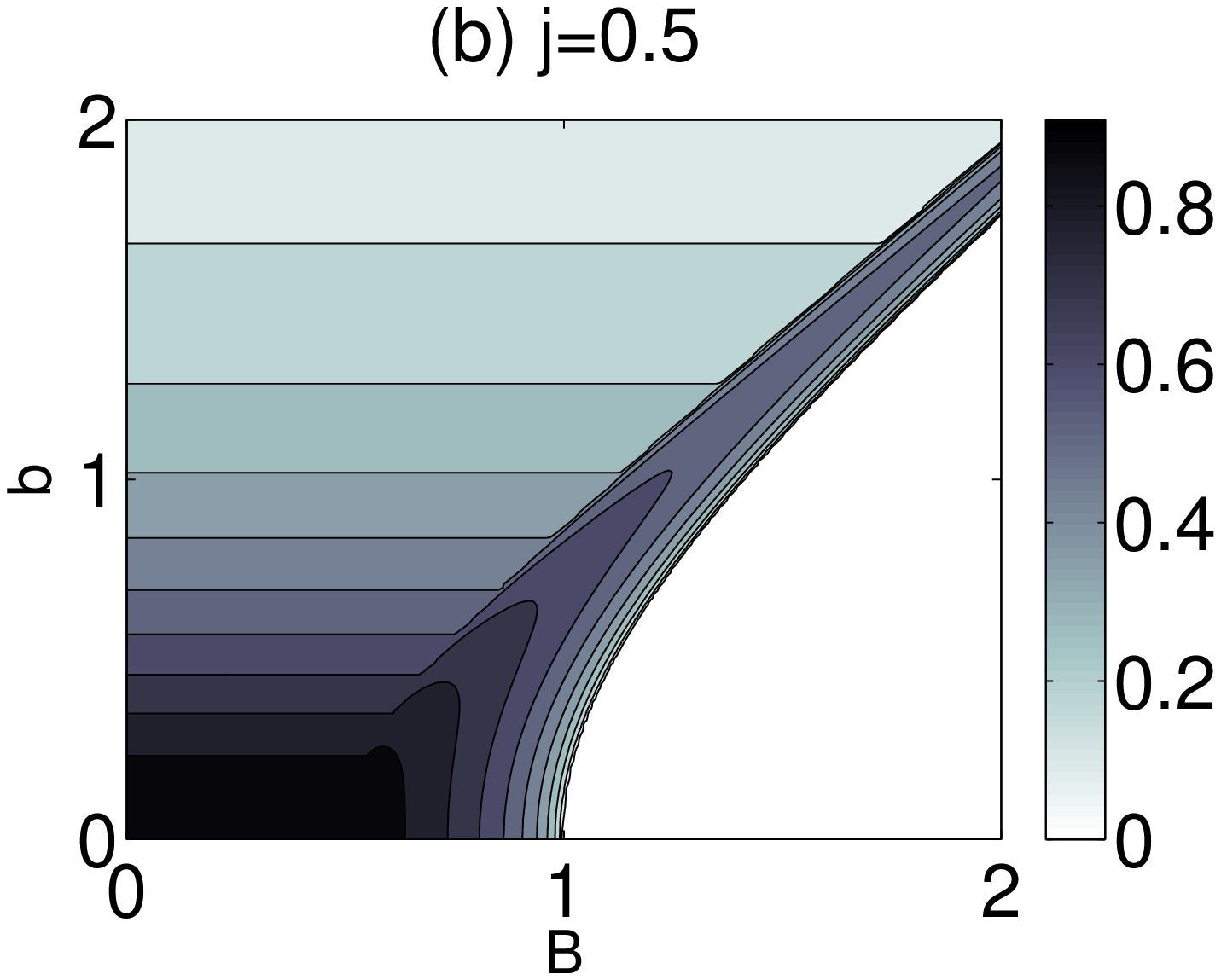} \\
  \includegraphics[width=13em, clip]{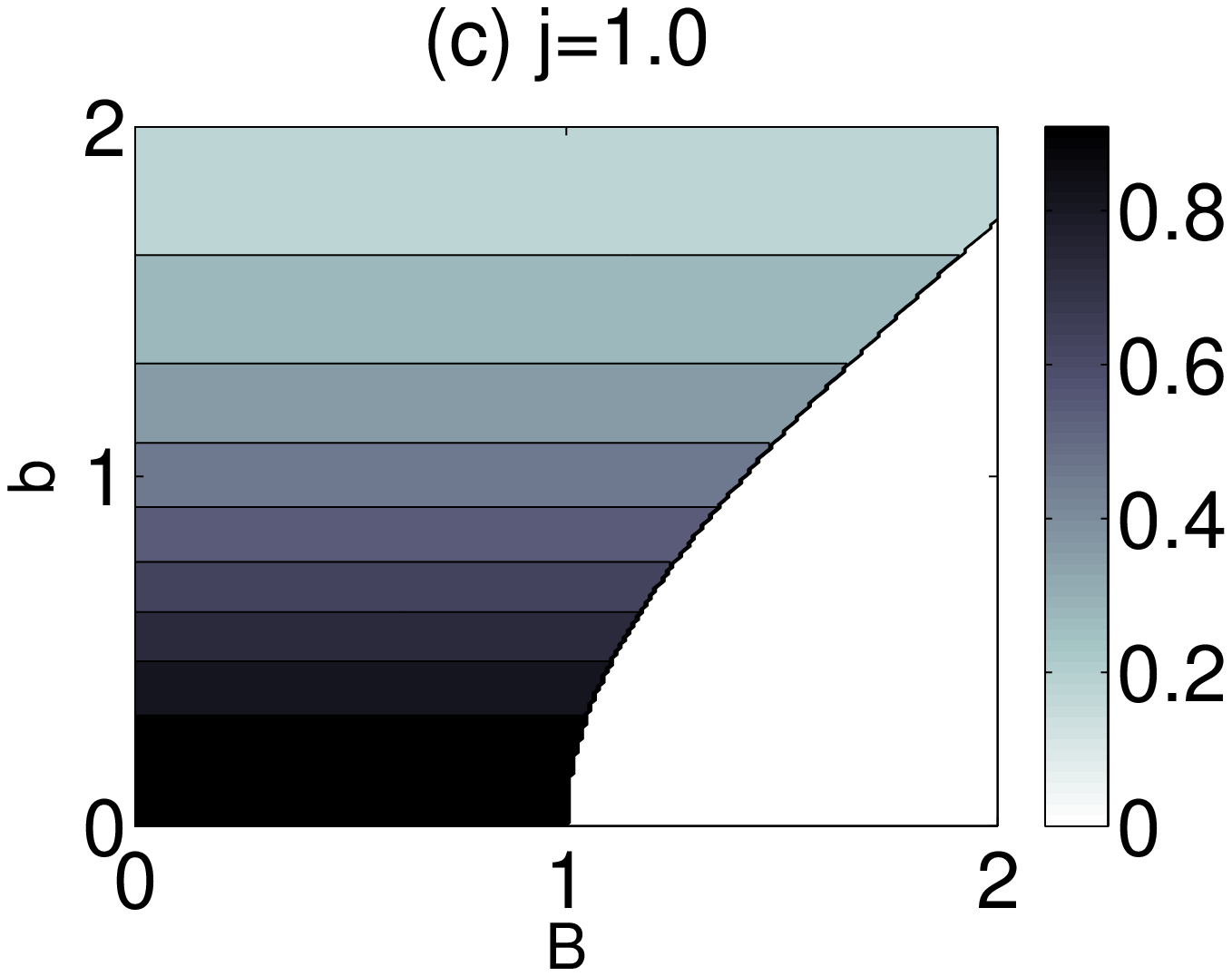} &
   \includegraphics[width=13em, clip]{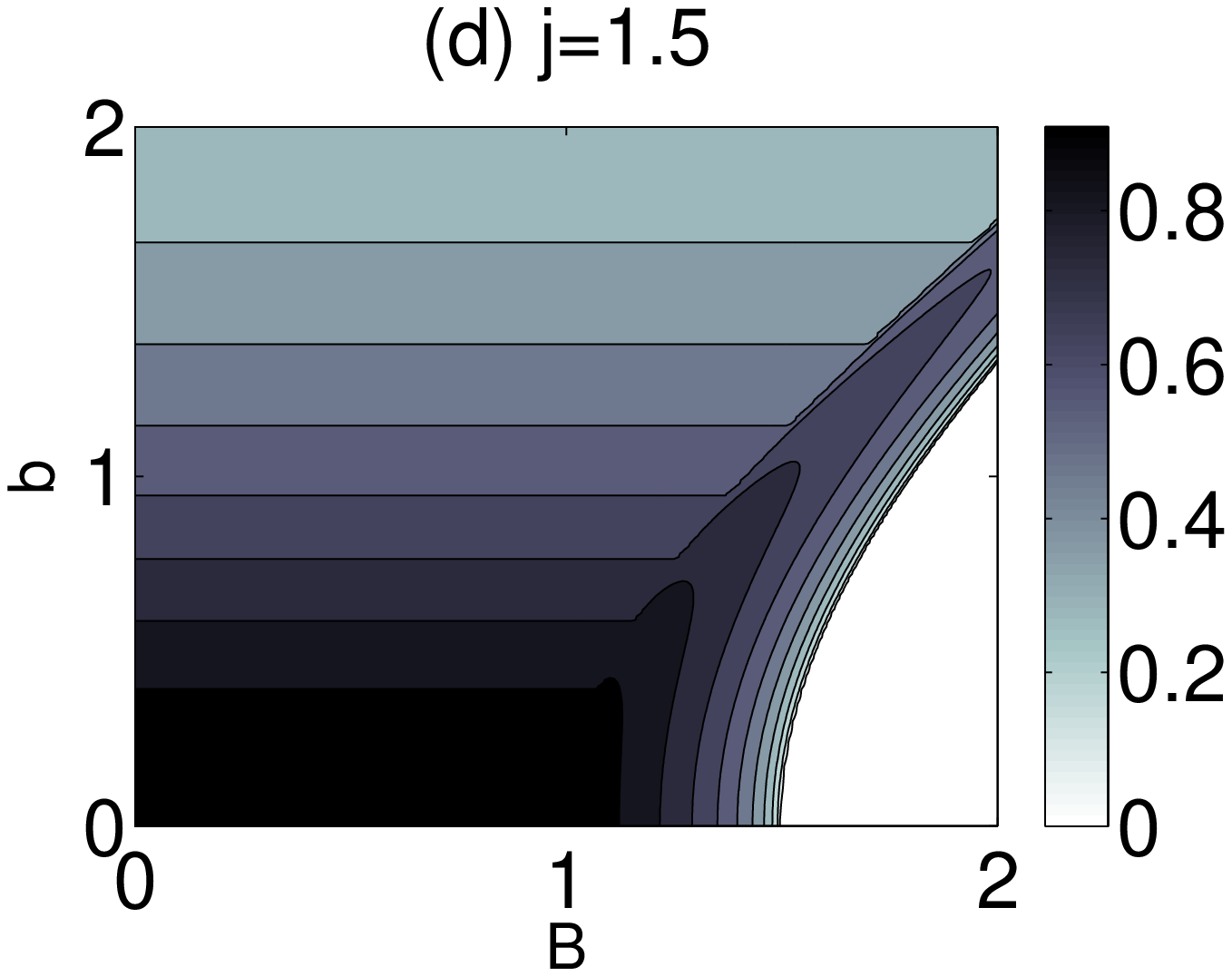}
  \end{tabular}
  \caption{
The Meyer-Wallach measure as a function of $B$ and $b$.
}
	\label{Fig:MeyerWallach2}
\end{figure}

\begin{figure}[tb]
\centering
\begin{tabular}{cc}
 \includegraphics[width=13em, clip]{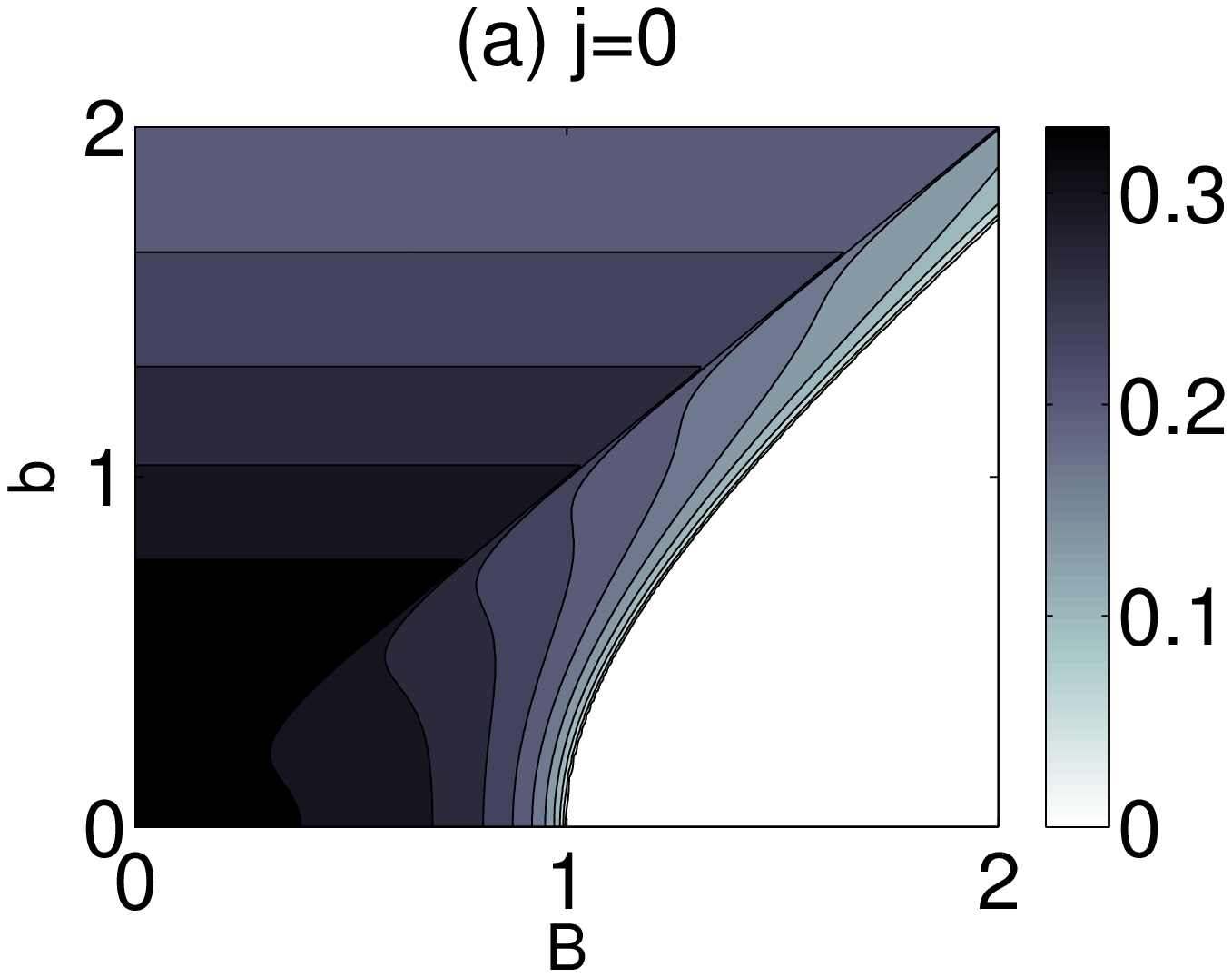} &
 \includegraphics[width=13em, clip]{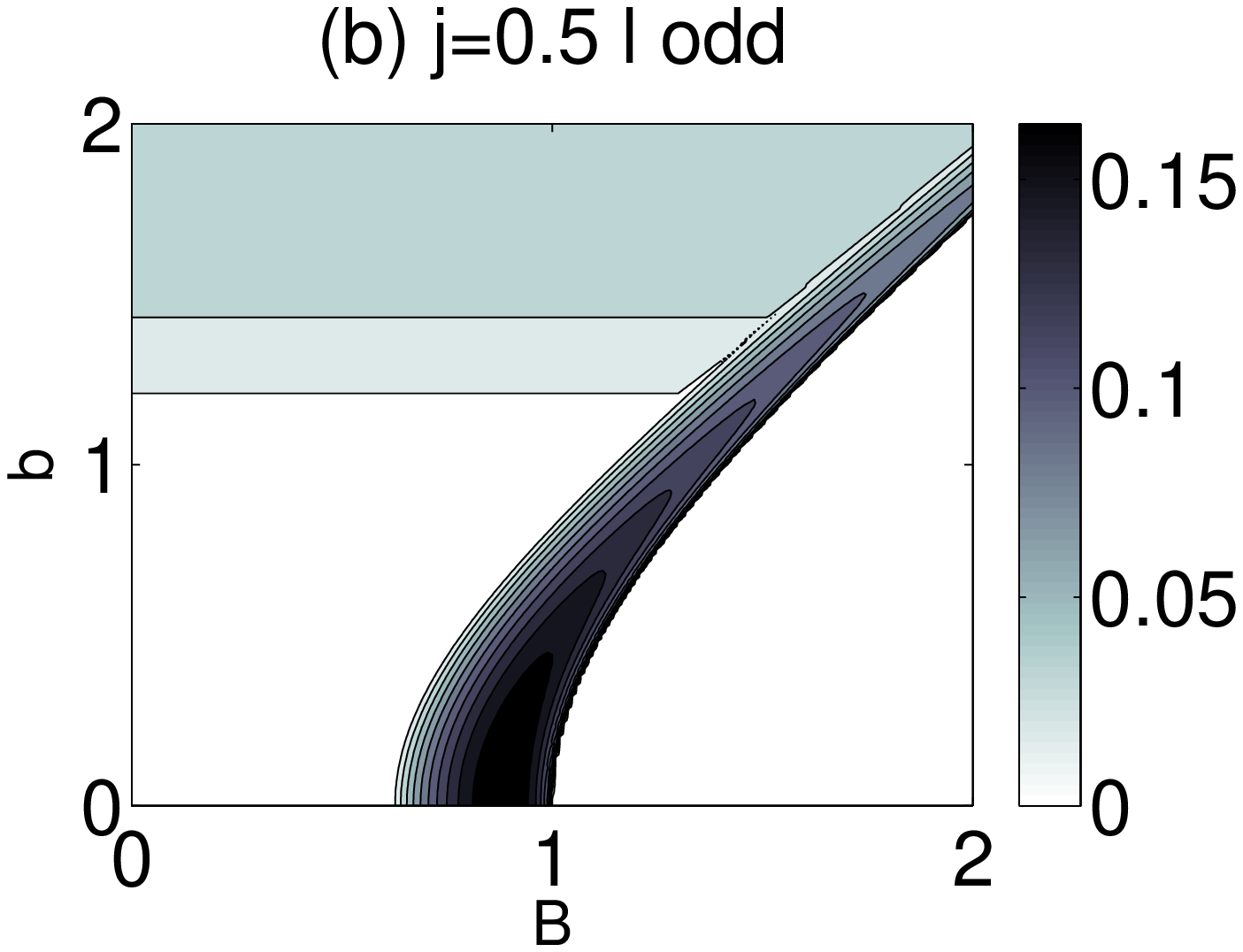} \\
  \includegraphics[width=13em, clip]{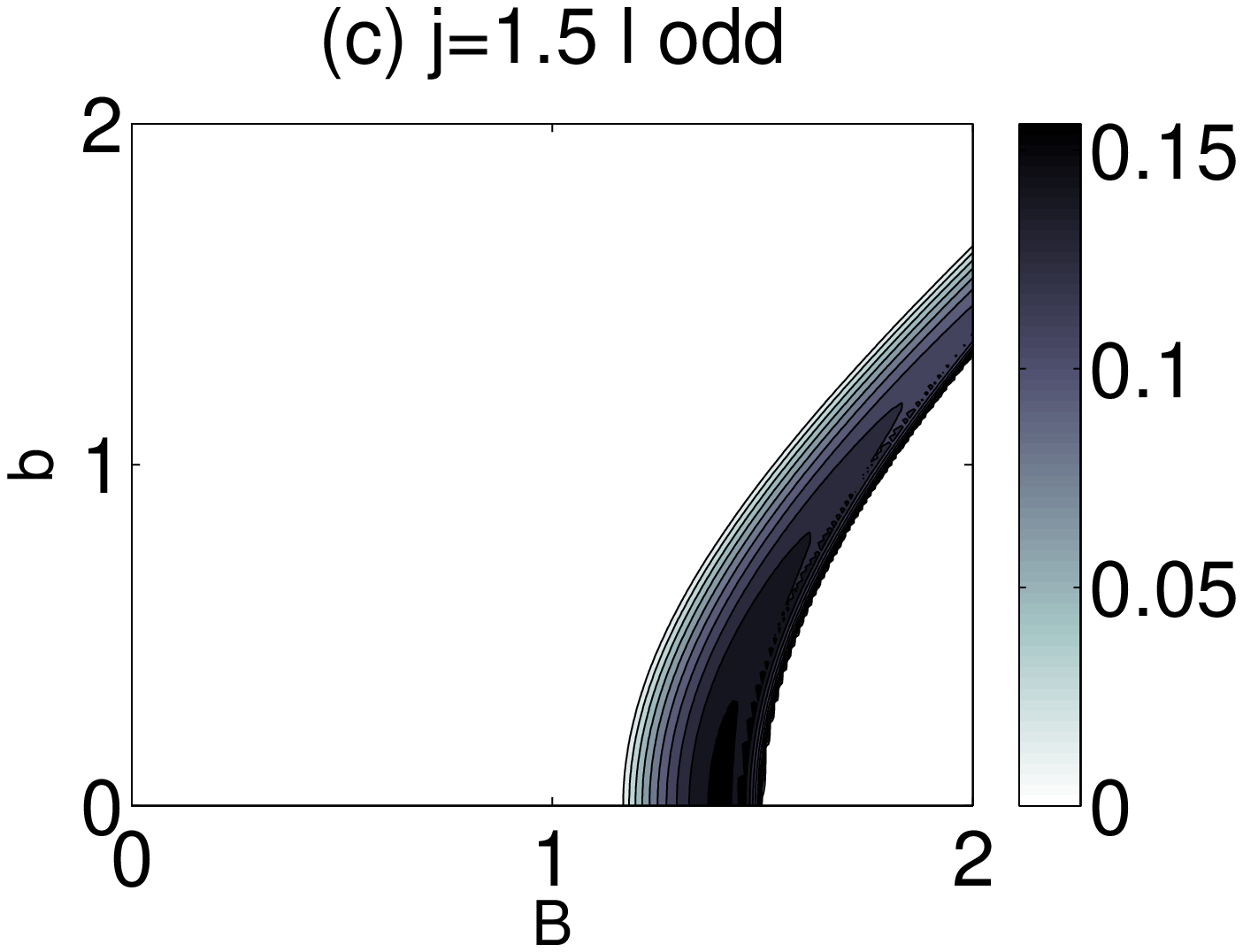} &
    \includegraphics[width=13em, clip]{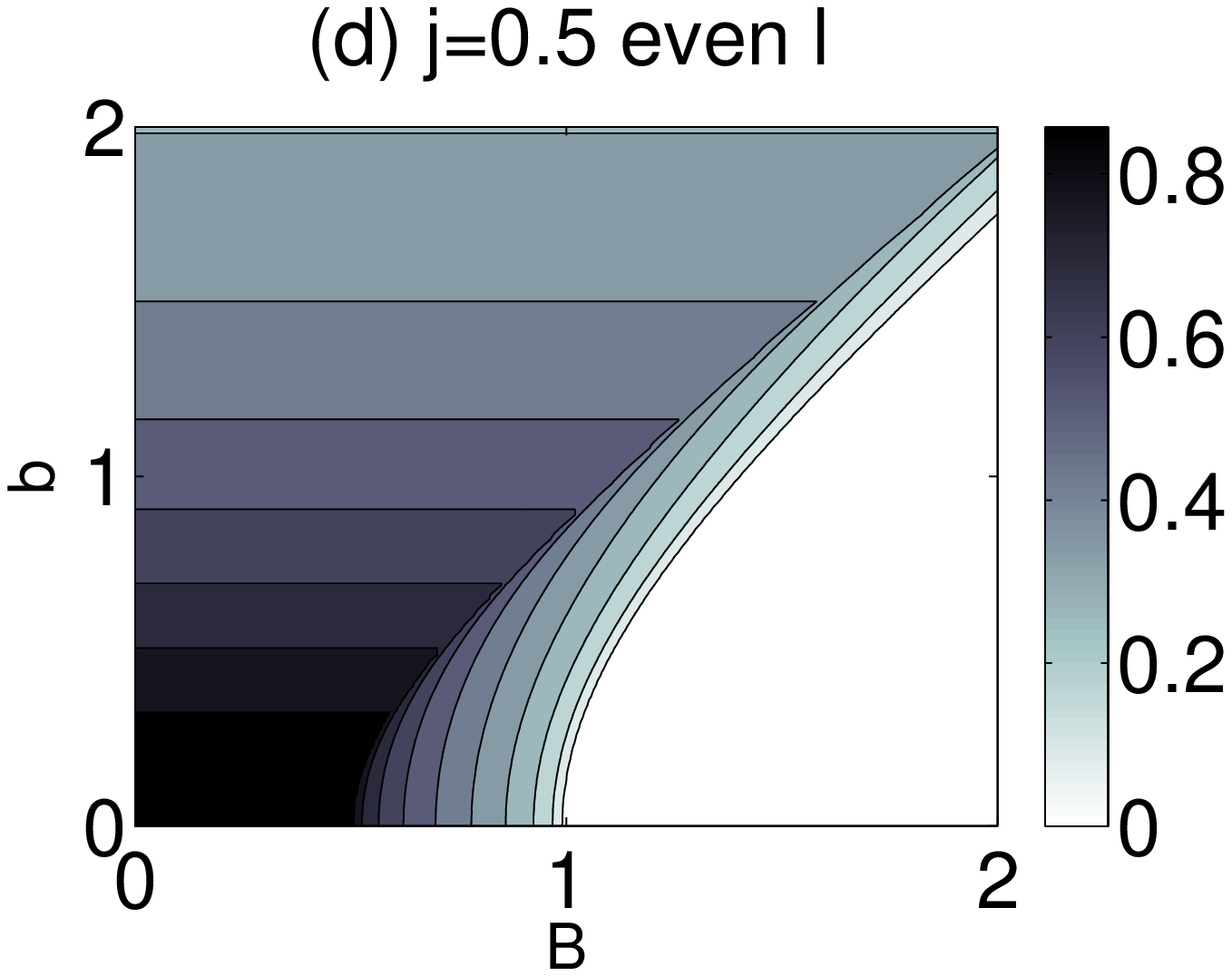} \\
   \includegraphics[width=13em, clip]{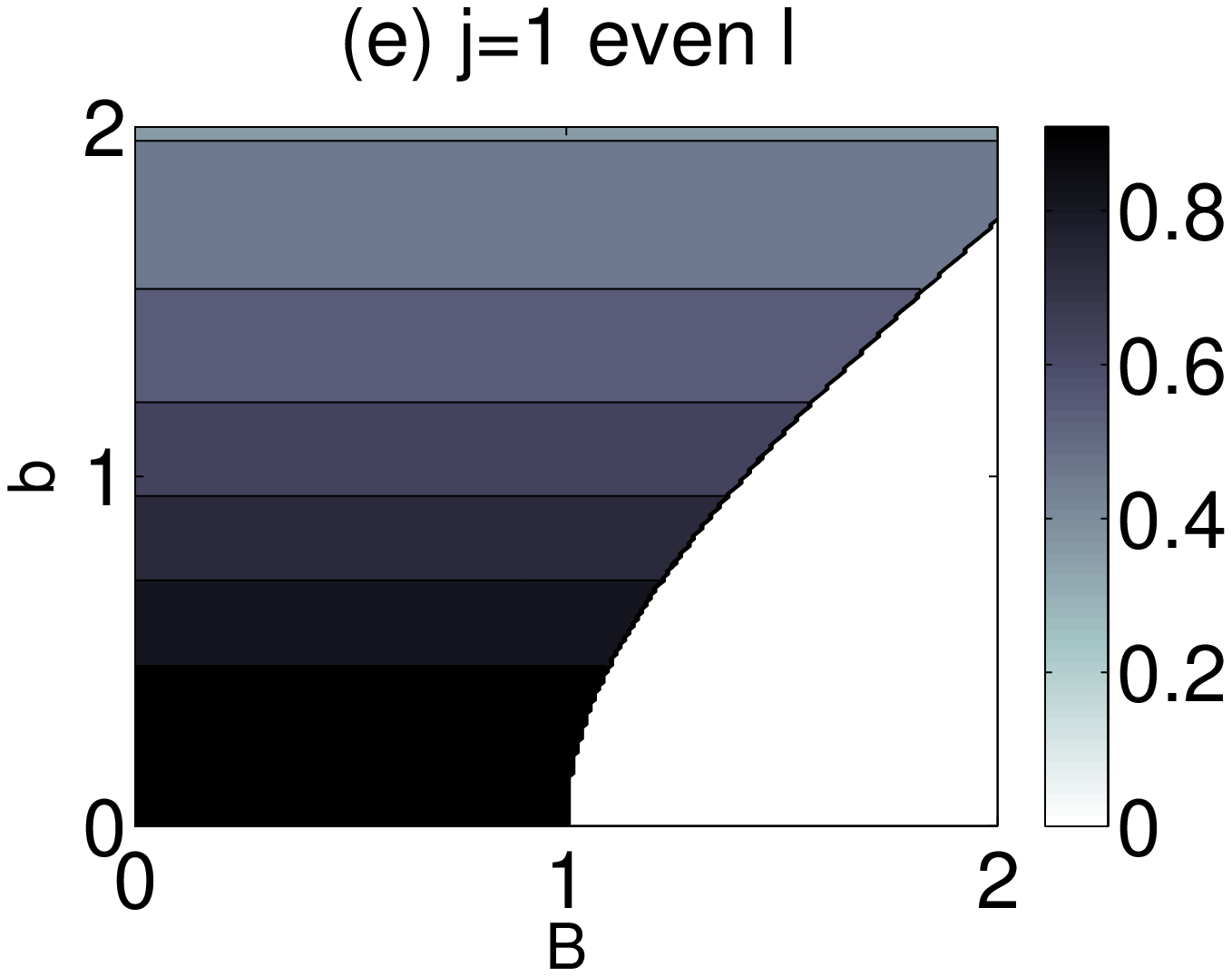} &
     \includegraphics[width=13em, clip]{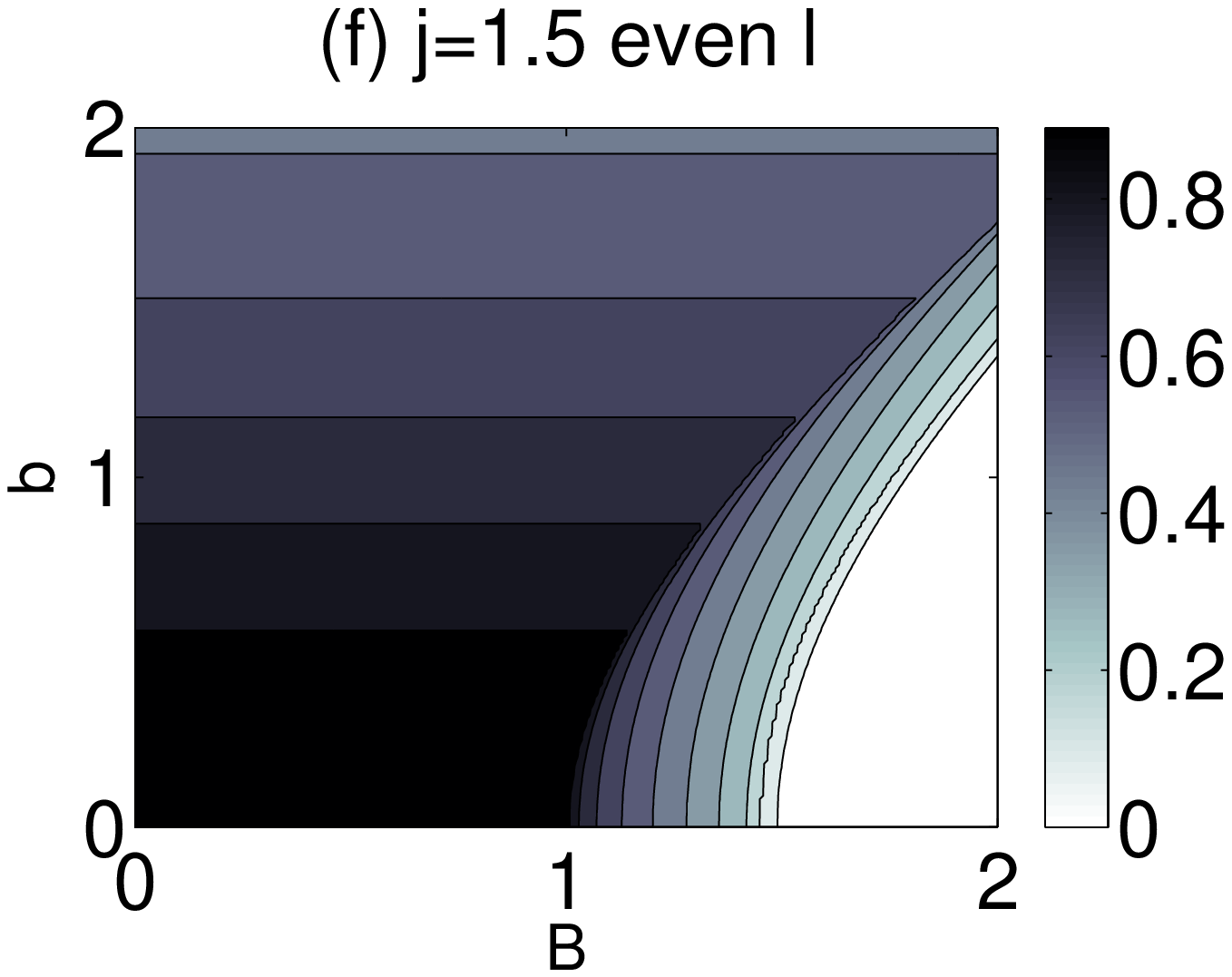}
  \end{tabular}
  \caption{
The nearest neighbour concurrence as a function of $B$ and $b$ as $T \rightarrow 0$.
When $j=0$, (a), the concurrence for odd and for even sites is identical.}
\label{Fig:NNConcT0j}
\end{figure}

 \begin{figure}[tb]
 \centering
 \begin{tabular}{cc}
 \includegraphics[width=13em, clip]{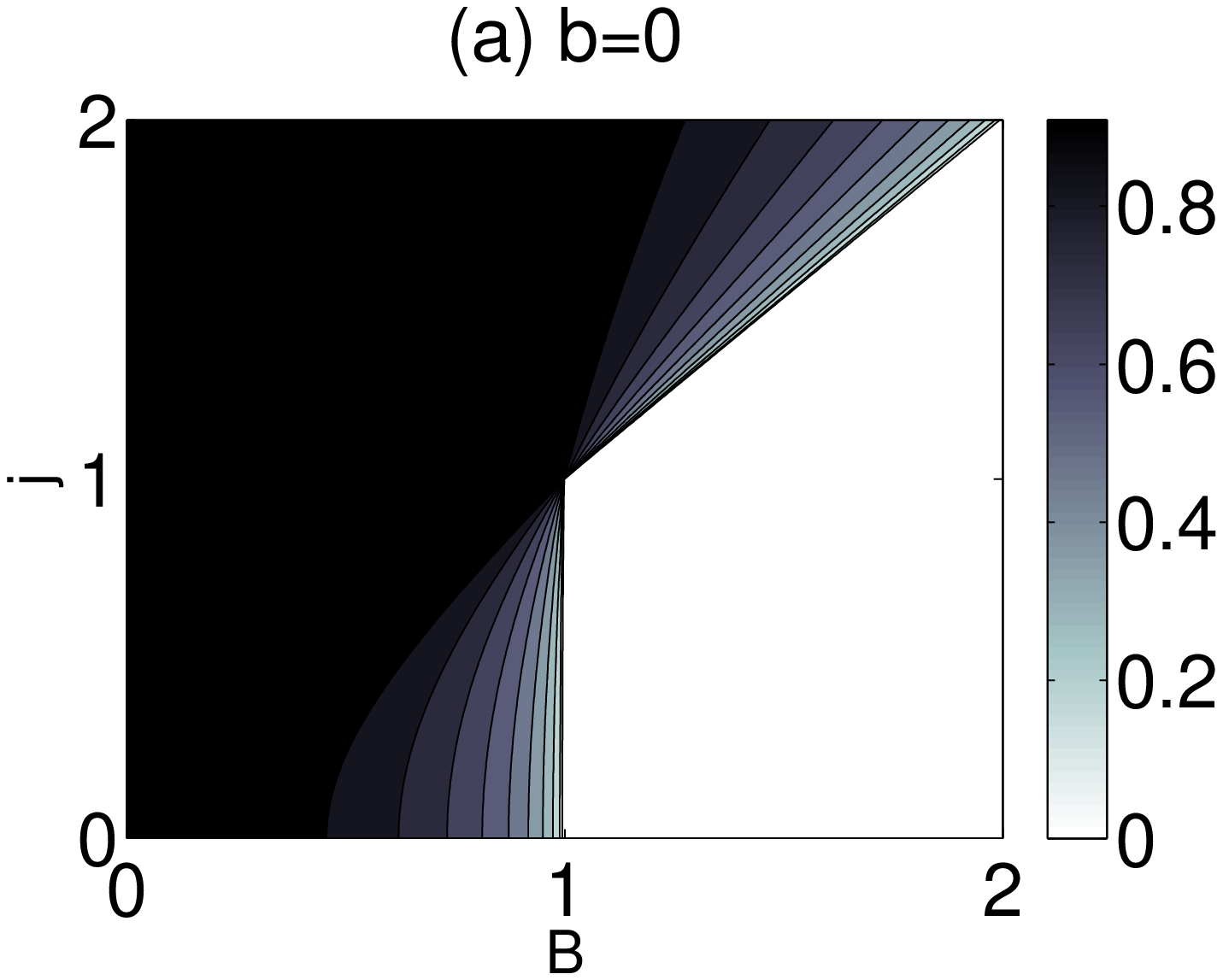} &
 \includegraphics[width=13em, clip]{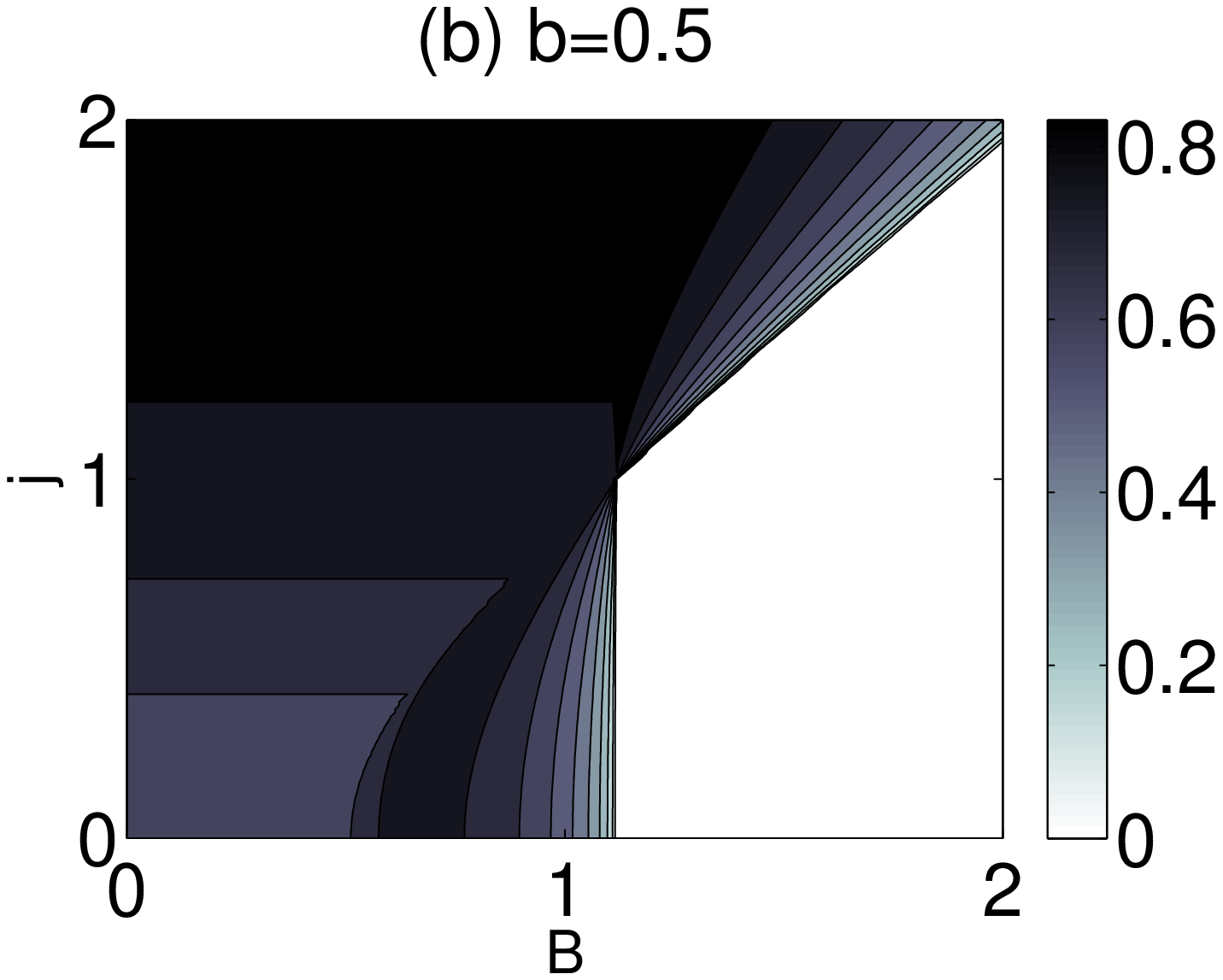} \\
   \includegraphics[width=13em, clip]{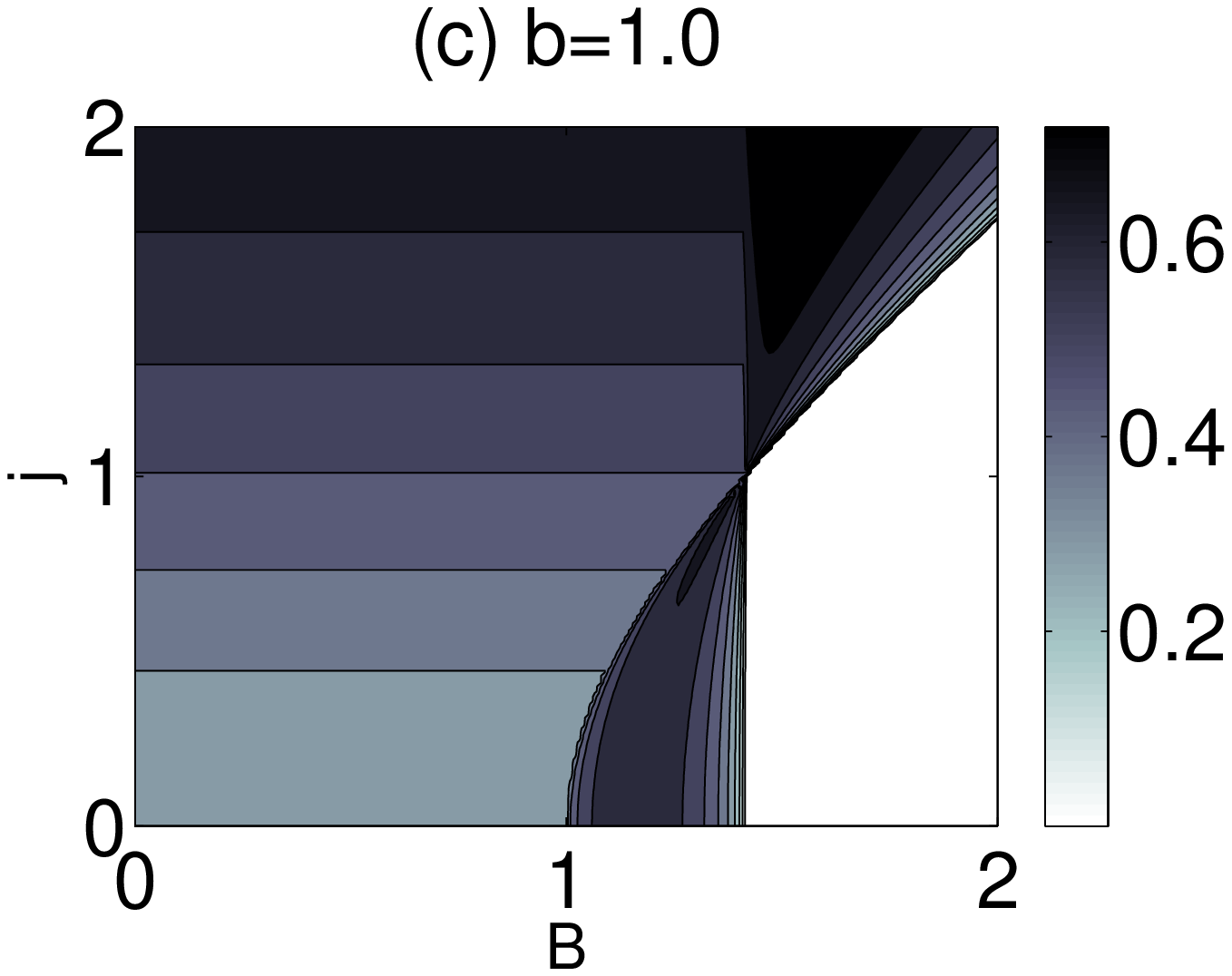} &
     \includegraphics[width=13em, clip]{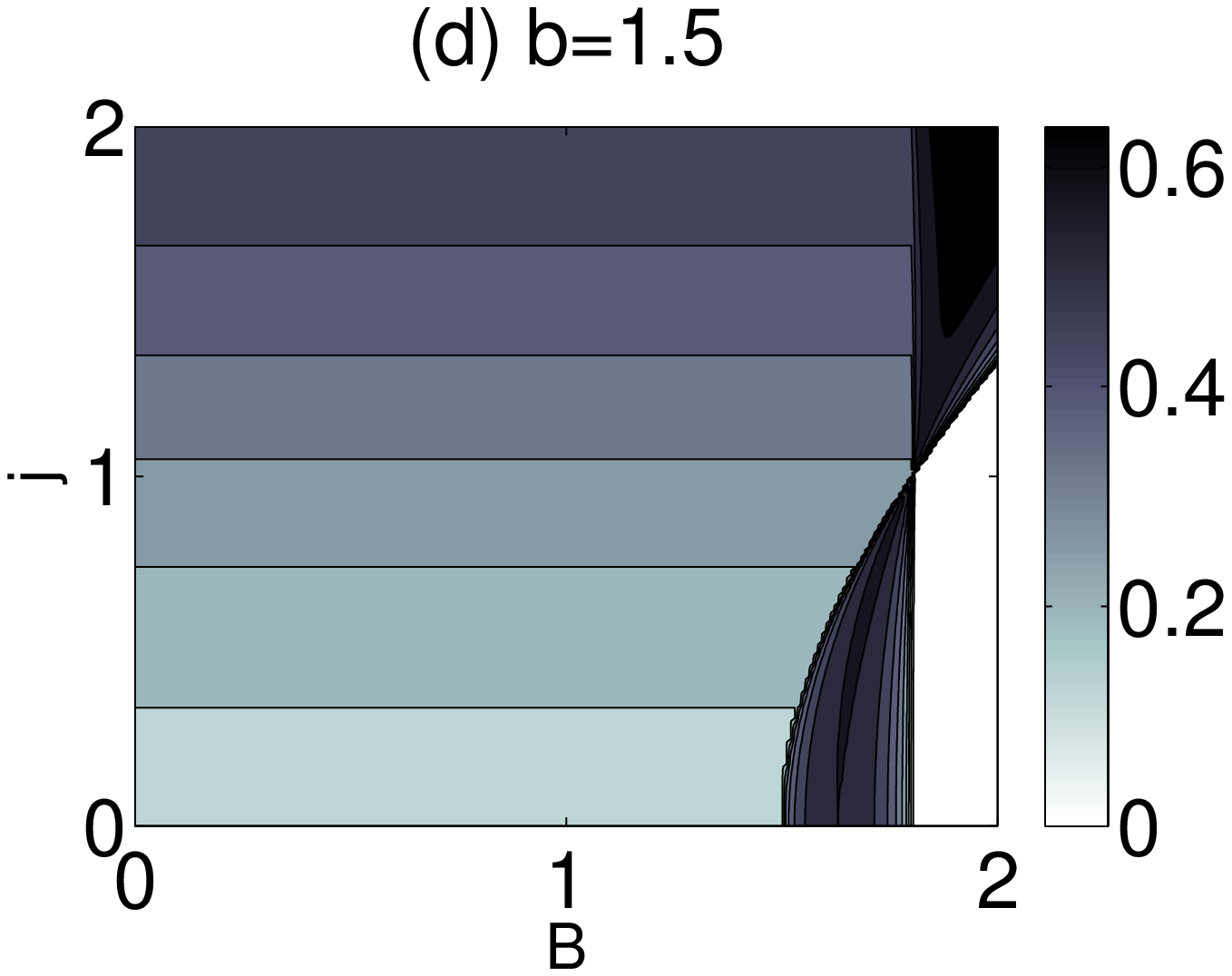}
  \end{tabular}
  \caption{
The Meyer-Wallach measure as a function of $B$ and $j$.}
	\label{Fig:MeyerWallach3}
\end{figure}

\begin{figure}[tb]
\centering
\begin{tabular}{cc}
 \includegraphics[width=13em, clip]{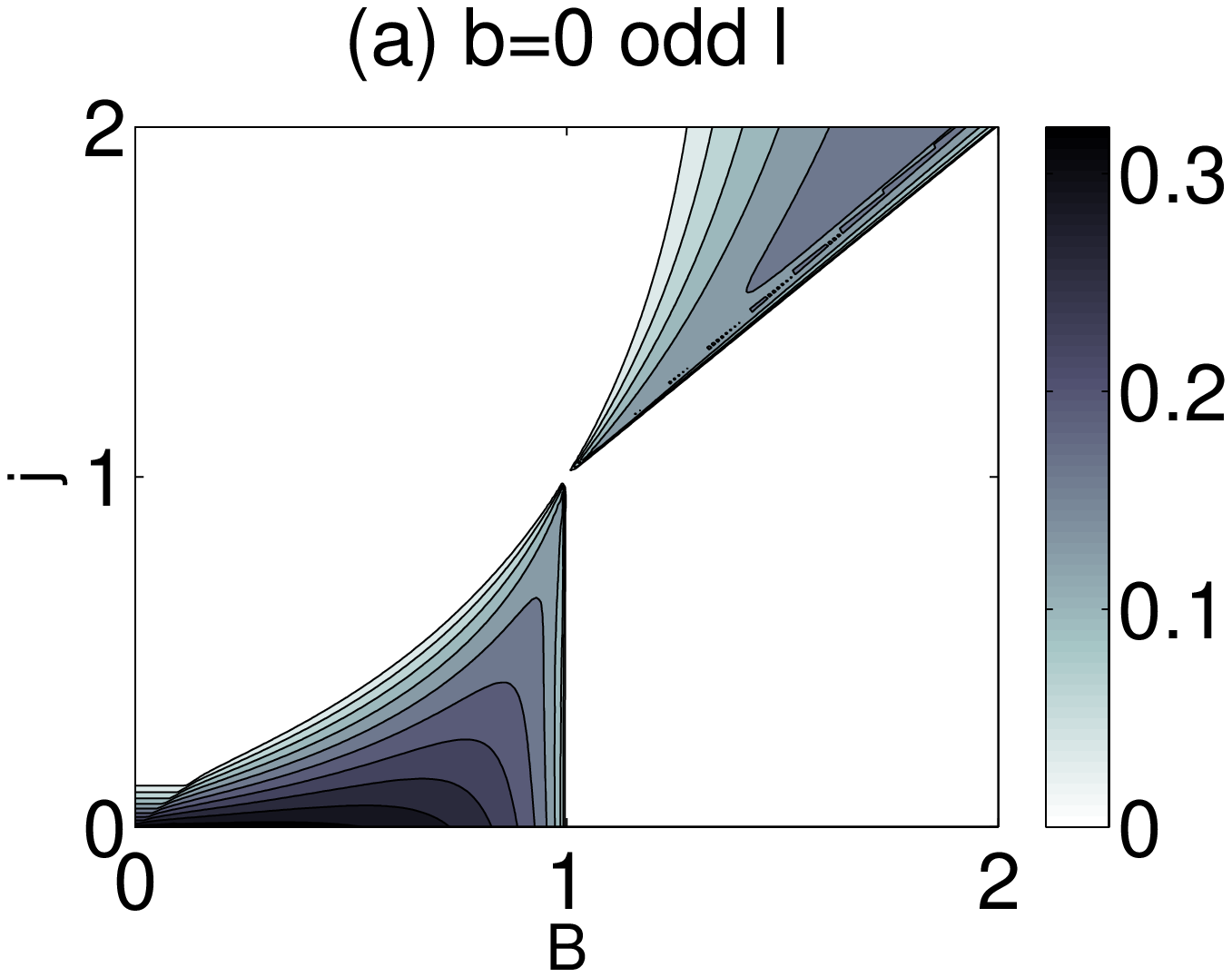} &
 \includegraphics[width=13em, clip]{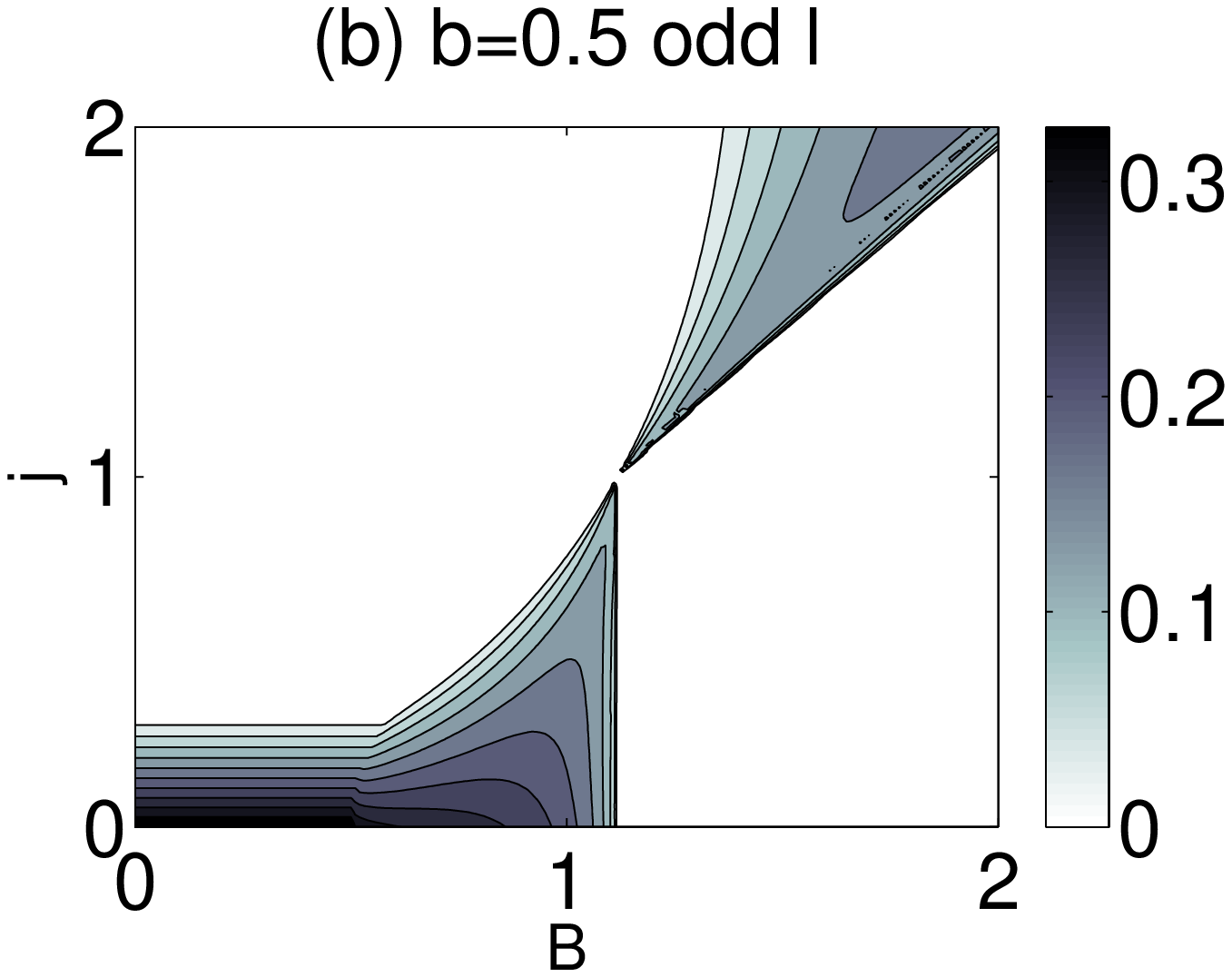} \\
  \includegraphics[width=13em, clip]{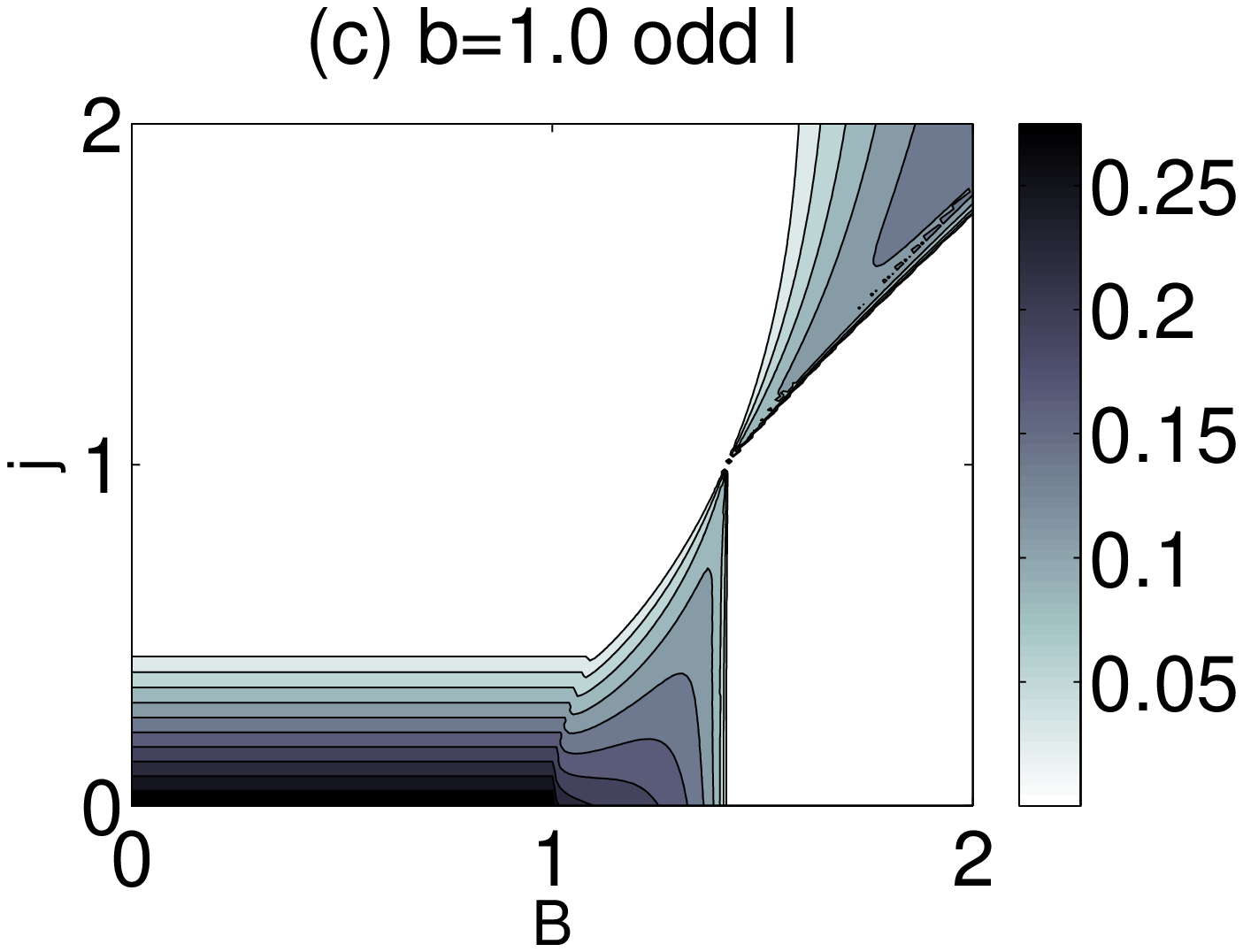} &
  \includegraphics[width=13em, clip]{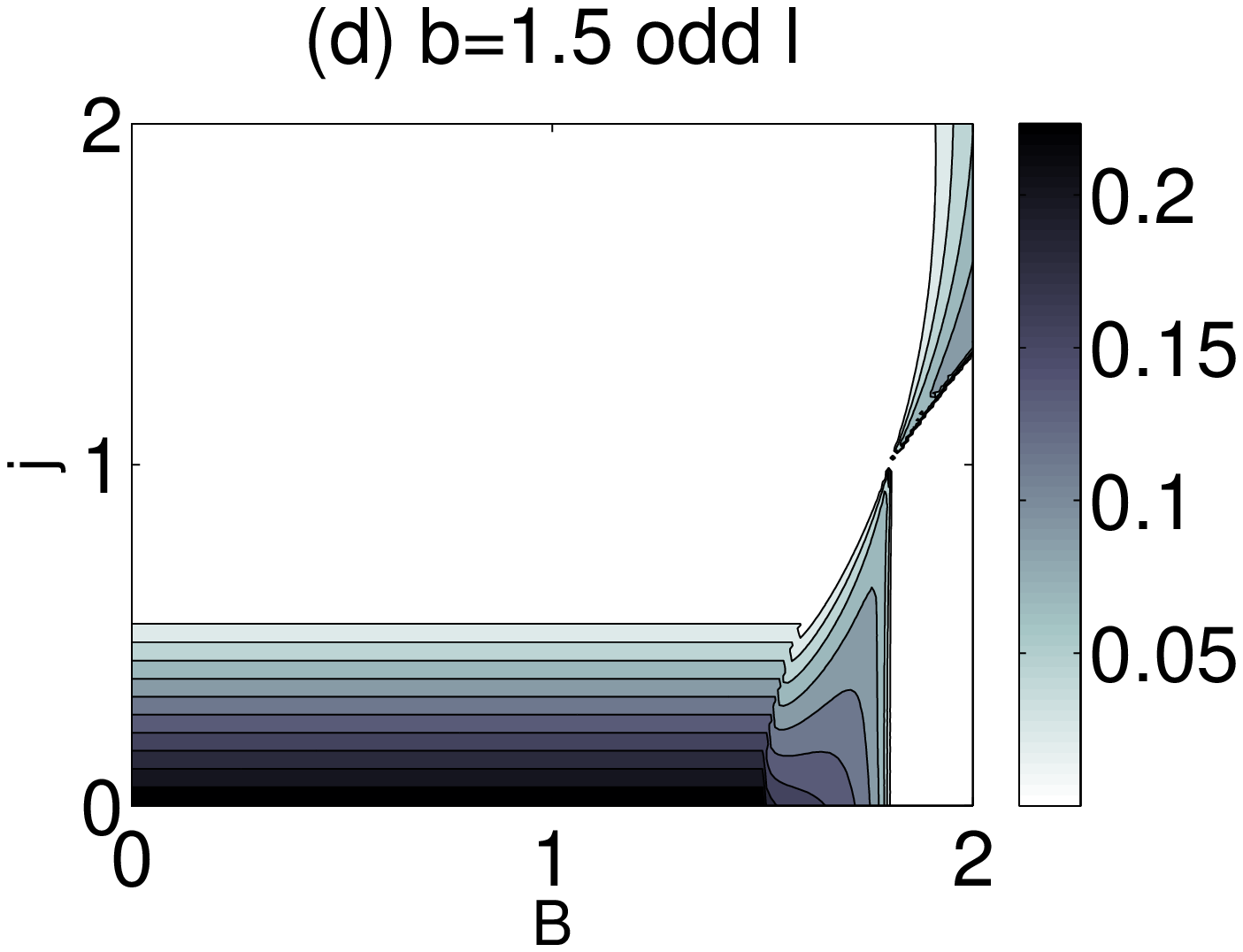} \\
  \includegraphics[width=13em, clip]{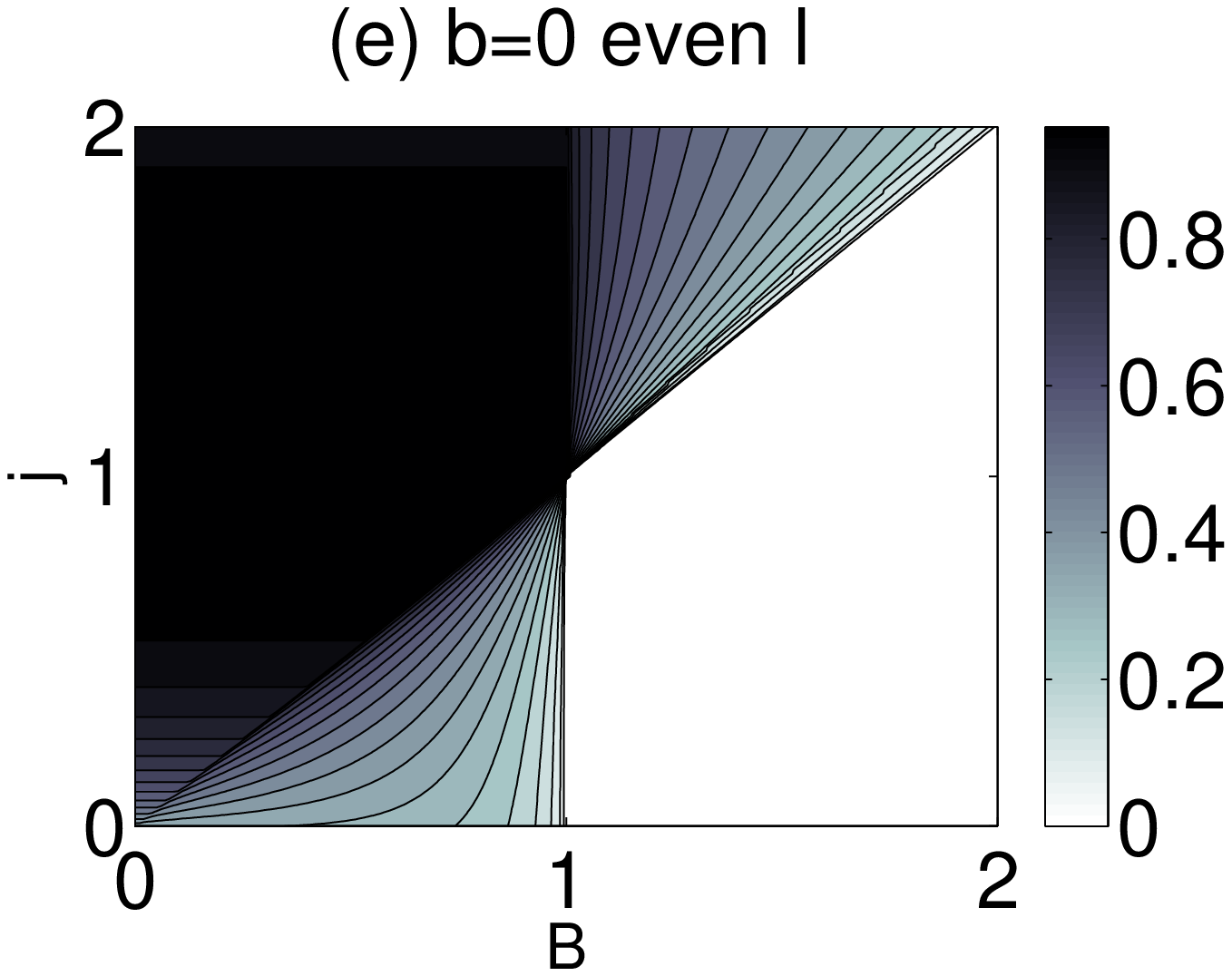} &
    \includegraphics[width=13em, clip]{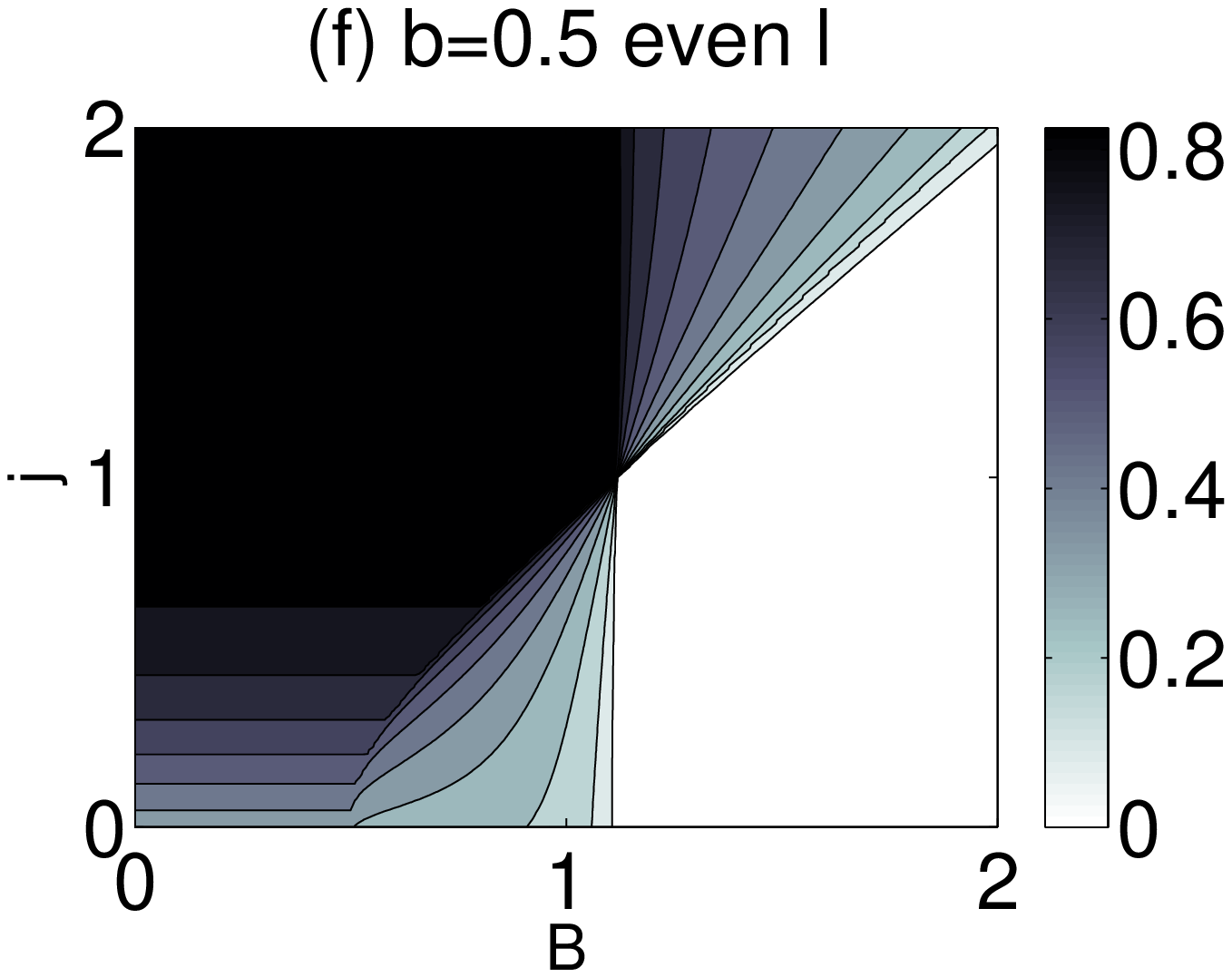} \\
   \includegraphics[width=13em, clip]{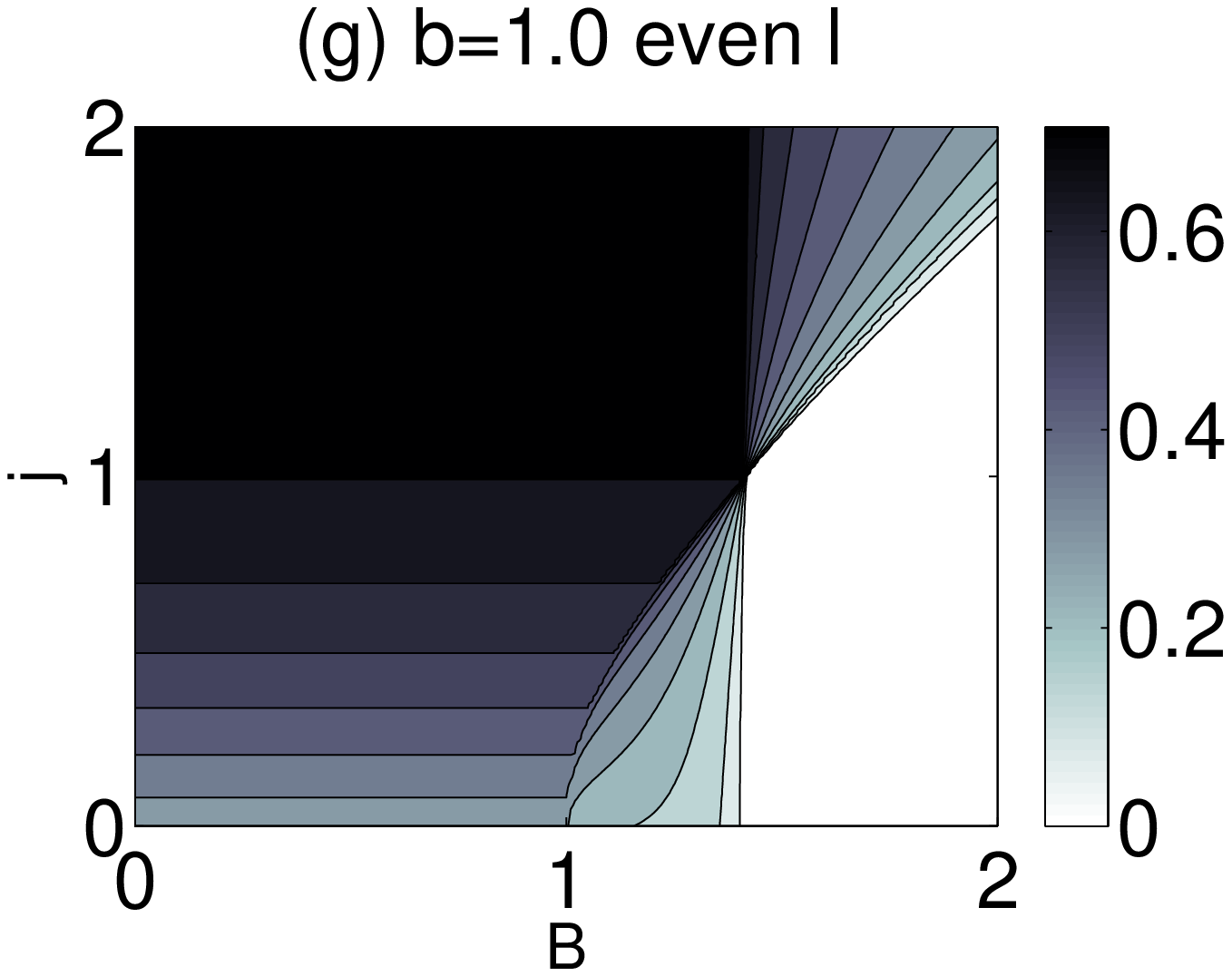} &
     \includegraphics[width=13em, clip]{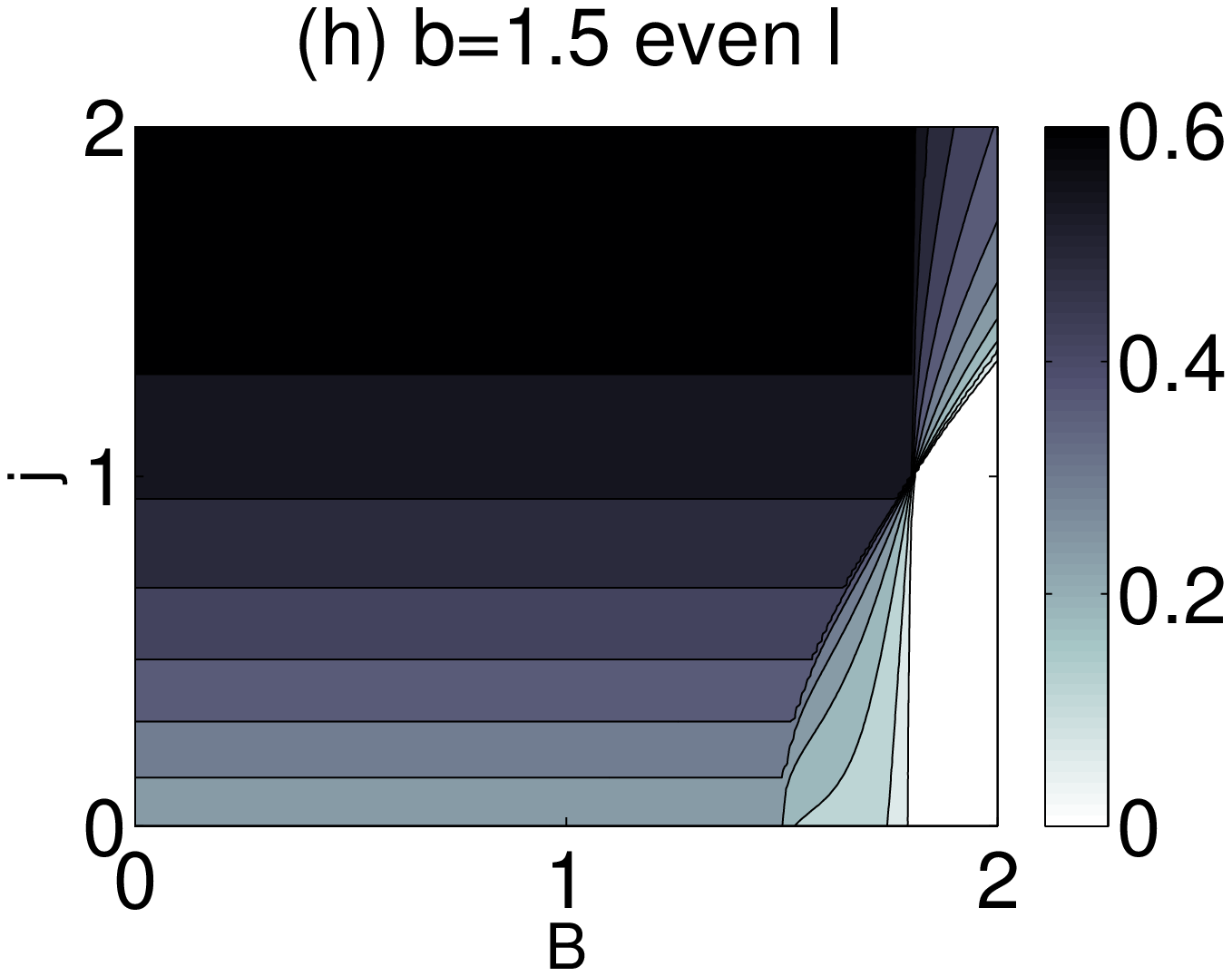} \\
     \includegraphics[width=13em, clip]{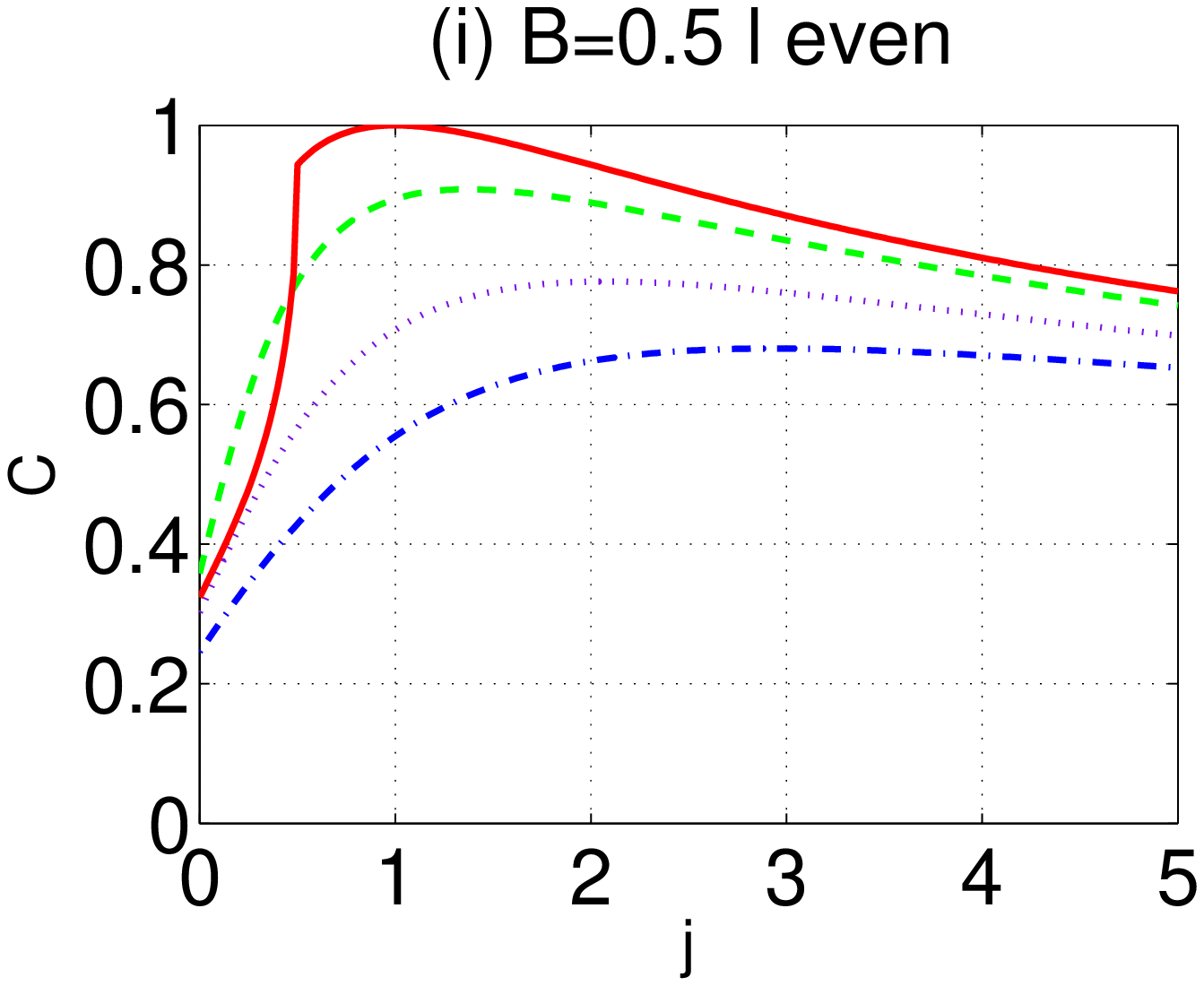}
  \end{tabular}
  \caption{
The nearest neighbour concurrence as a function of $B$ and $j$ as $T \rightarrow 0$.
(i) plots the concurrence against $j$ for $l$ even, $B=0.5$ to show that the concurrence
decreases for high enough $j$ for $b=0$ (solid red line), $b=0.5$ (dashed green line),
$b=1$ (dotted purple line) and $b=1.5$ (dot-dashed blue line).}
\label{Fig:NNConcT0b}
\end{figure}

Next, we discuss the effects of $b$ and $j$ on the Meyer-Wallach measure and the concurrence.

First, we consider the effect of $b$ on the entanglement, shown in Figs.~\ref{Fig:MeyerWallach2}
and~\ref{Fig:NNConcT0j}.
How the concurrence behaves in the presence of the alternating fields is highly dependent on whether
the site is odd or even. In general, due to the larger coupling from an even
site to an odd site, both nearest (NN) and next nearest neighbour (NNN) concurrence
is higher, with the opposite being true from an odd
to an even site. Entanglement for the Meyer-Wallach measure and odd site NN concurrence
remains large only when $B$ is between the two QPTs, i.e. $\sqrt{j^2+b^2}<B <\sqrt{J^2+b^2}$
when $j<J$ or $\sqrt{J^2+b^2}< B <\sqrt{j^2+b^2}$ when $J<j$. For both measures, the
maximum entanglement is at $B \sim b$ for large magnetic fields.
These results are understandable from the fact that a large magnetic field aligns spins in the
same direction leading to a separable state. When $b$ and $B$ are large, entanglement can be
large only for $B \sim b$ since the magnetic field on odd sites is canceled in such cases.

The even site NN concurrence does not follow this pattern, and instead a larger
amount of entanglement tends to be present when $B<\sqrt{j^2+b^2}$ for $j<J$ or when $B<\sqrt{J^2+b^2}$
for $J<j$.

On the other hand, the Meyer-Wallach measure and the concurrence vary differently with the alternating
coupling constant $j$ as shown in Figs.~\ref{Fig:MeyerWallach3} and \ref{Fig:NNConcT0b} where
the entanglement is plotted as a function of $j$.
The Meyer-Wallach measure is an increasing function with $j$ except in the vicinity of the QCPs
while the concurrence is in general a decreasing function of $j$. Fig.~\ref{Fig:NNConcT0b} (i)
shows that for each $b$, there is a non-zero value of $j$ for which the concurrence is the
maximum possible.
Since the Meyer-Wallach measure is a measure of the entanglement shared in the whole spin chain
and the concurrence measures the entanglement between two spins,
we can conclude from these results that as the alternating coupling constant $j$ increases,
the amount of entanglement shared amongst all spins increases.
Such global entanglement is not locally detected, in the sense that the entanglement
of the reduced two spin state is small.

The concurrence is closely related to the entanglement of formation.
In Fig.~\ref{Fig:NNConcT0b}, for odd sites, when $B>\sqrt{J^2+b^2}$,
increasing $j$ can increase concurrence, and for even sites, when $B<\sqrt{j^2+b^2}$,
increasing $j$ increases the concurrence until a maximal value is reached, after which the
concurrence decreases again. Thus a low but non-zero value of $j$ can be beneficial to
the extraction of maximally entangled state.

Another interesting feature common to both the Meyer-Wallach measure and the concurrence, demonstrated in
Figs.~\ref{Fig:MeyerWallach2}-~\ref{Fig:NNConcT0b}, is that below the region of the QCPs, the entanglement
is constant as $B$ varies. That is, the magnetic field $B$ is not a dominant parameter for entanglement below
the QCPs. The QPTs are often intuitively understood as occurring due to the balance between the strength of
the coupling constants and that of the magnetic fields. As such, the dominant parameters of this
system are the coupling constants (the magnetic fields) below (above) the QCPs in general.
Our results support this intuition from the viewpoint of entanglement in the sense that the magnetic 
field $B$ does not change the entanglement below the QCPs. On the other hand, the entanglement is 
sensitive to the change of the alternating magnetic field $b$ even below the QCPs, which 
demonstrates the difference between $B$ and $b$.

Finally, the NNN concurrence is reduced compared to NN concurrence as expected, but remains
reasonably high, especially for even sites where a non-zero value of $b$ increases the
entanglement. Further, increasing $b$ allows a spin chain
with larger values of $B$ to be entangled.

%%%%%%%%%%%%%%%%%%%%%%%%%%%%%%%%%%%%%%%%%%%%%%%%%%%%%%%%%%%%%%%%%%%%%%%%%%%%%%%%%%%%%%%
\subsection{Finite temperature}
%%%%%%%%%%%%%%%%%%%%%%%%%%%%%%%%%%%%%%%%%%%%%%%%%%%%%%%%%%%%%%%%%%%%%%%%%%%%%%%%%%%%%%%
\label{SubSec:Conc}

\begin{figure}[t]
\centering
\begin{tabular}{cc}
\epsfig{file=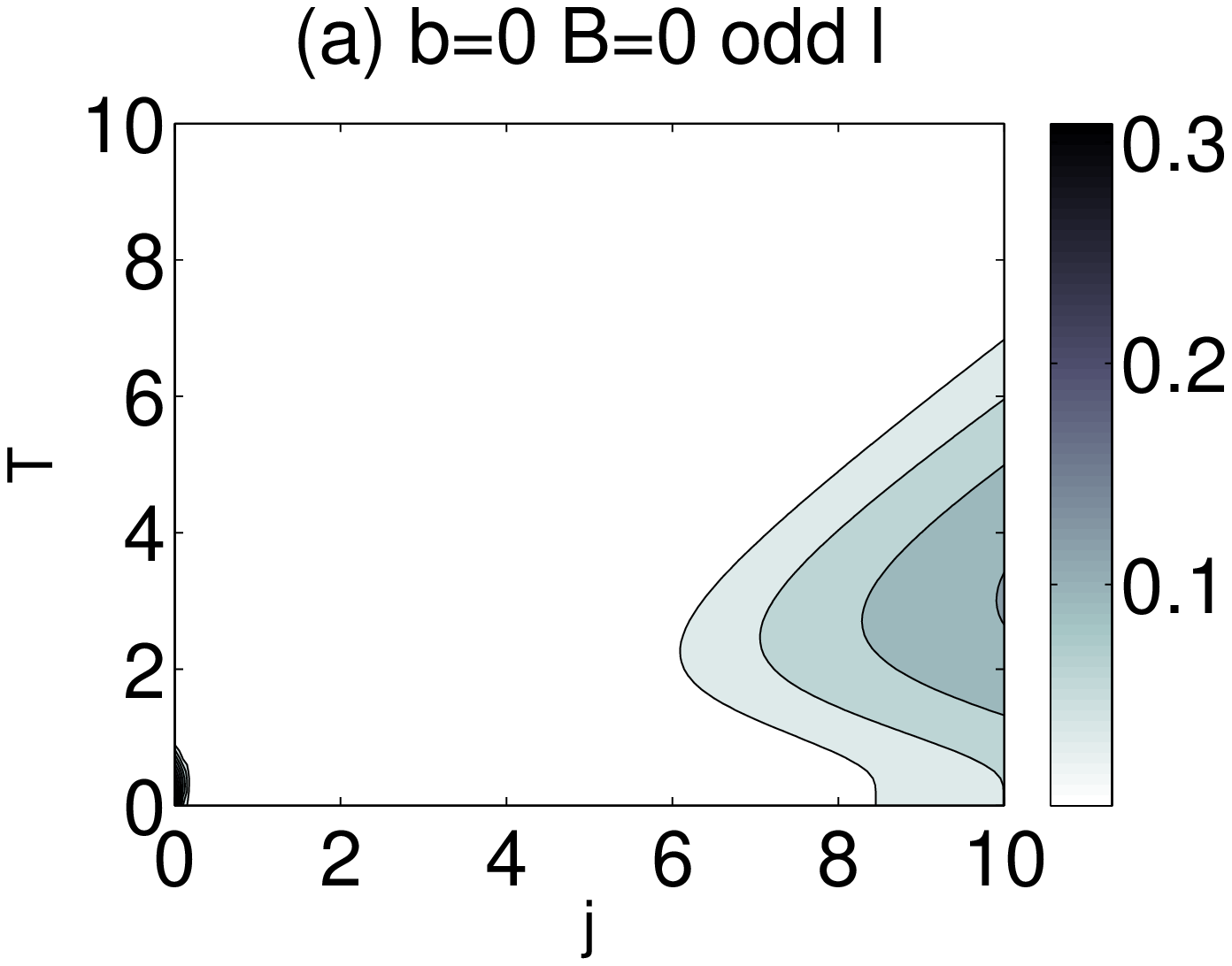,width=0.5\columnwidth,clip=} &
\epsfig{file=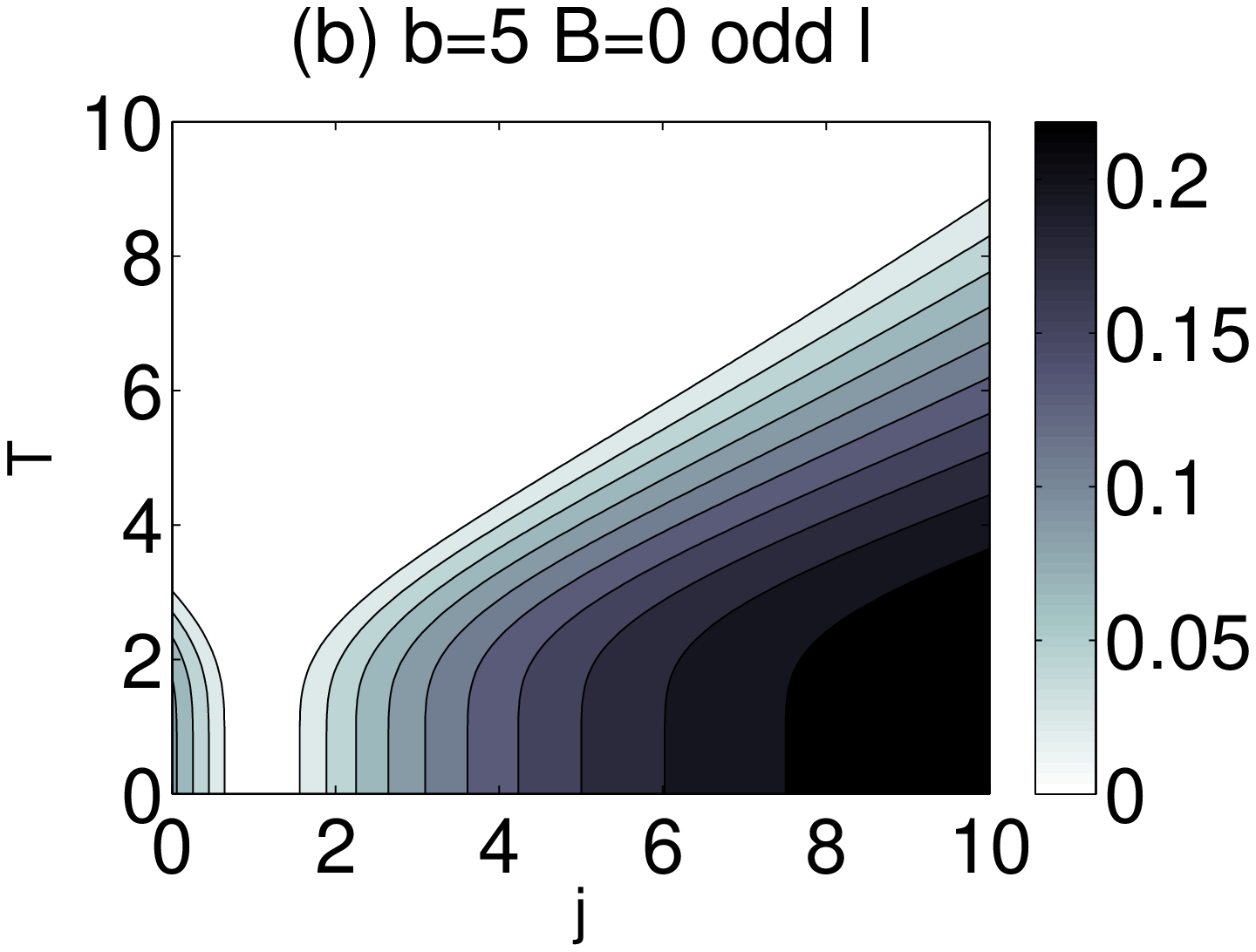,width=0.5\columnwidth,clip=} \\
\epsfig{file=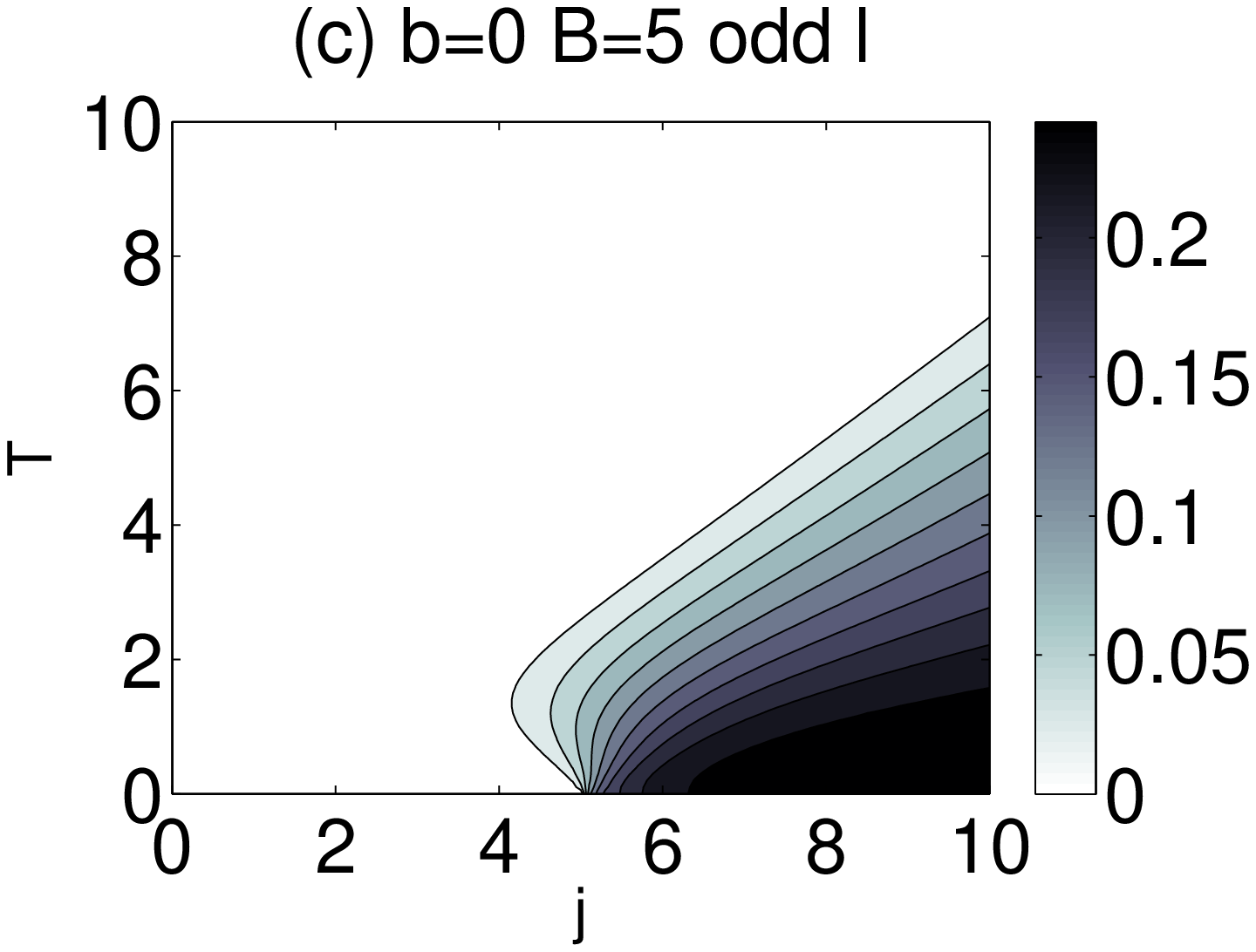,width=0.5\columnwidth,clip=} &
\epsfig{file=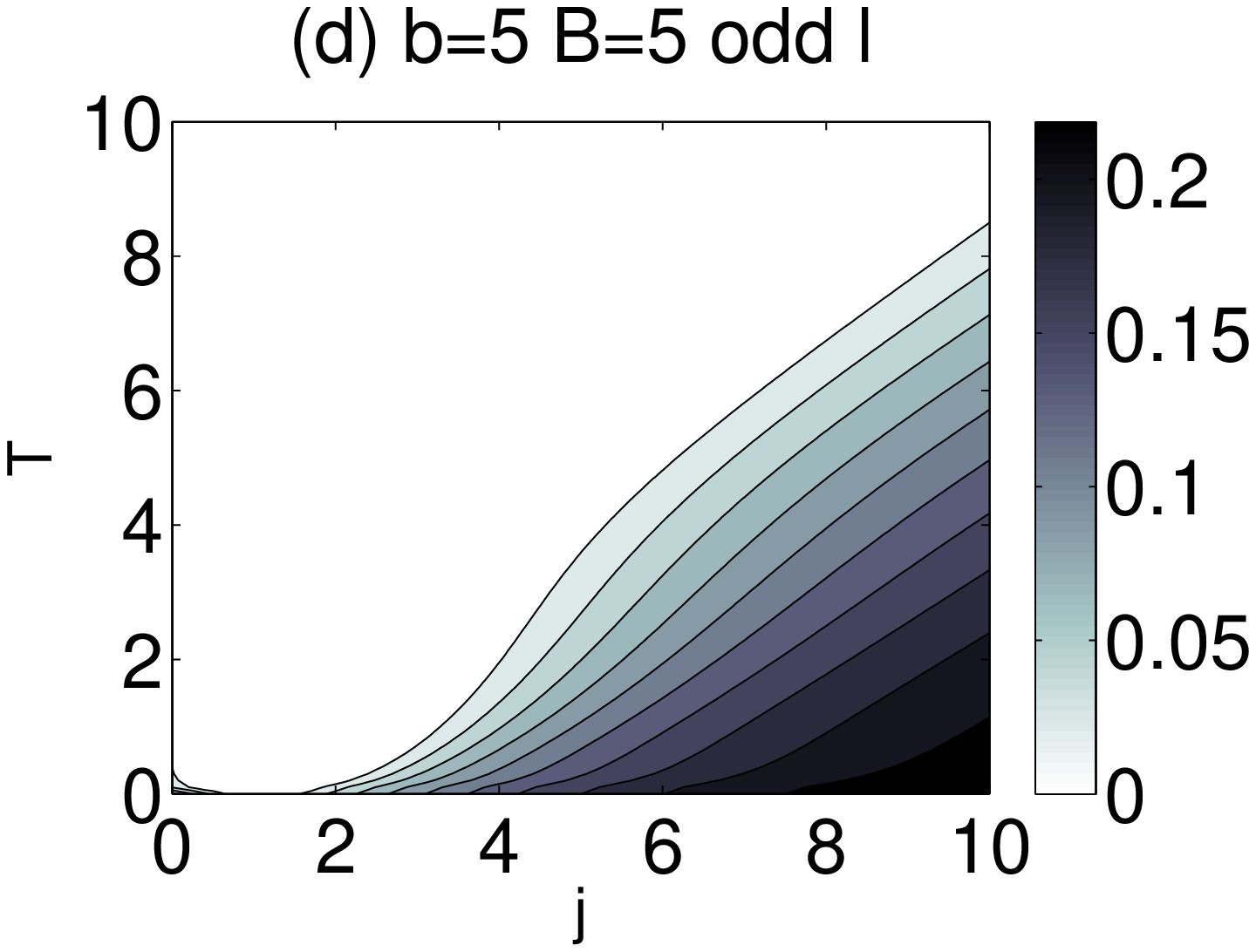,width=0.5\columnwidth,clip=} \\
\epsfig{file=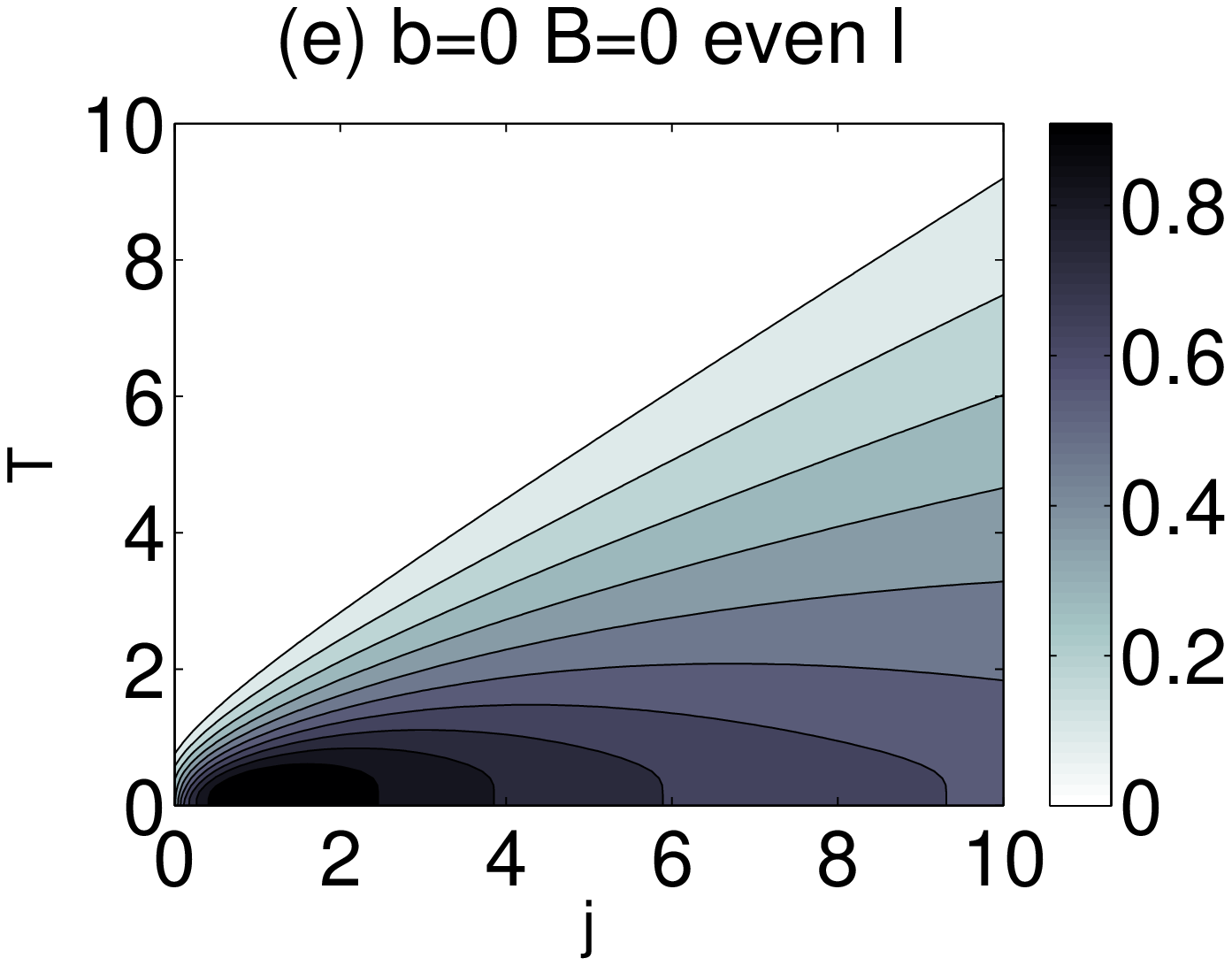,width=0.5\columnwidth,clip=} &
\epsfig{file=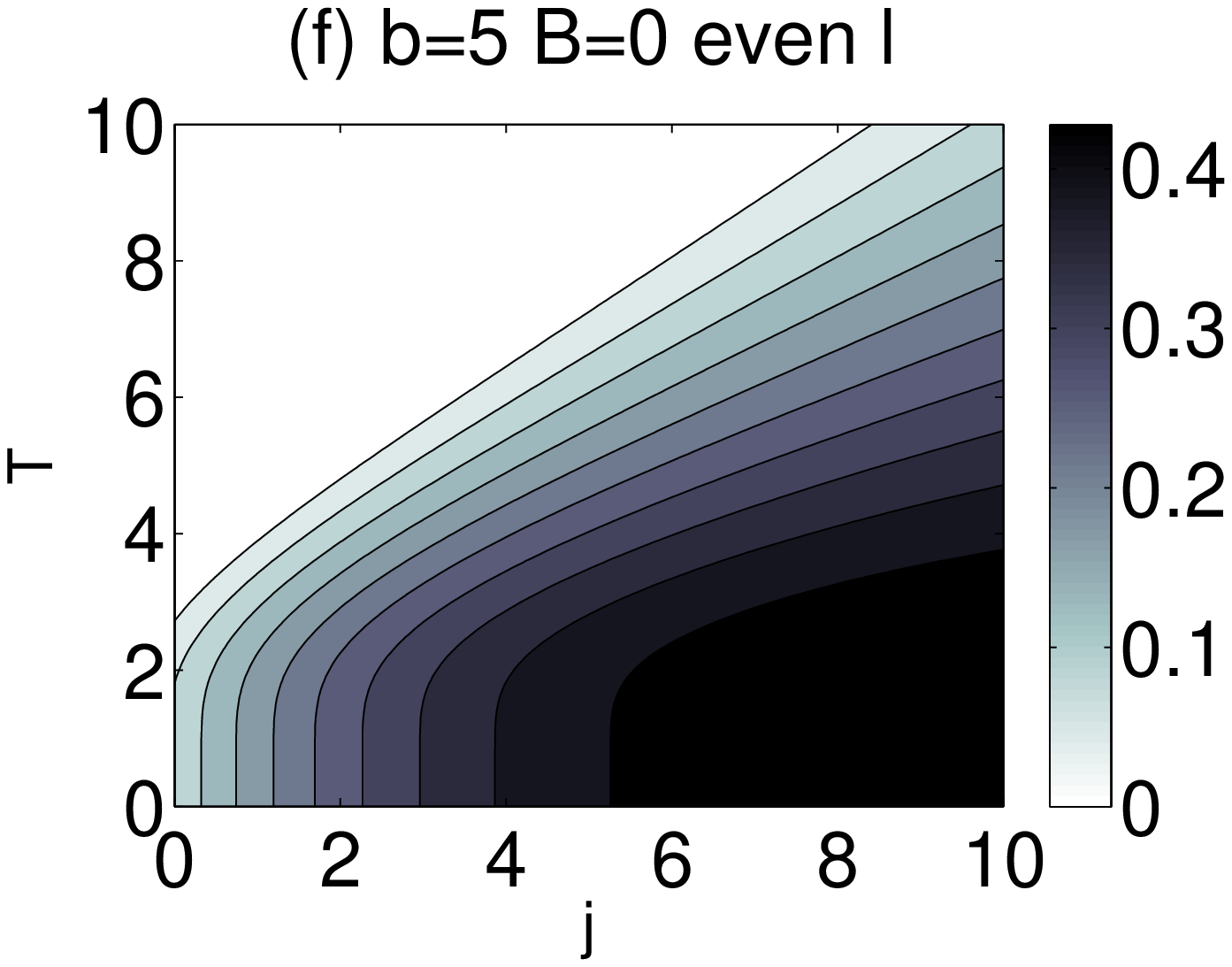,width=0.5\columnwidth,clip=} \\
\epsfig{file=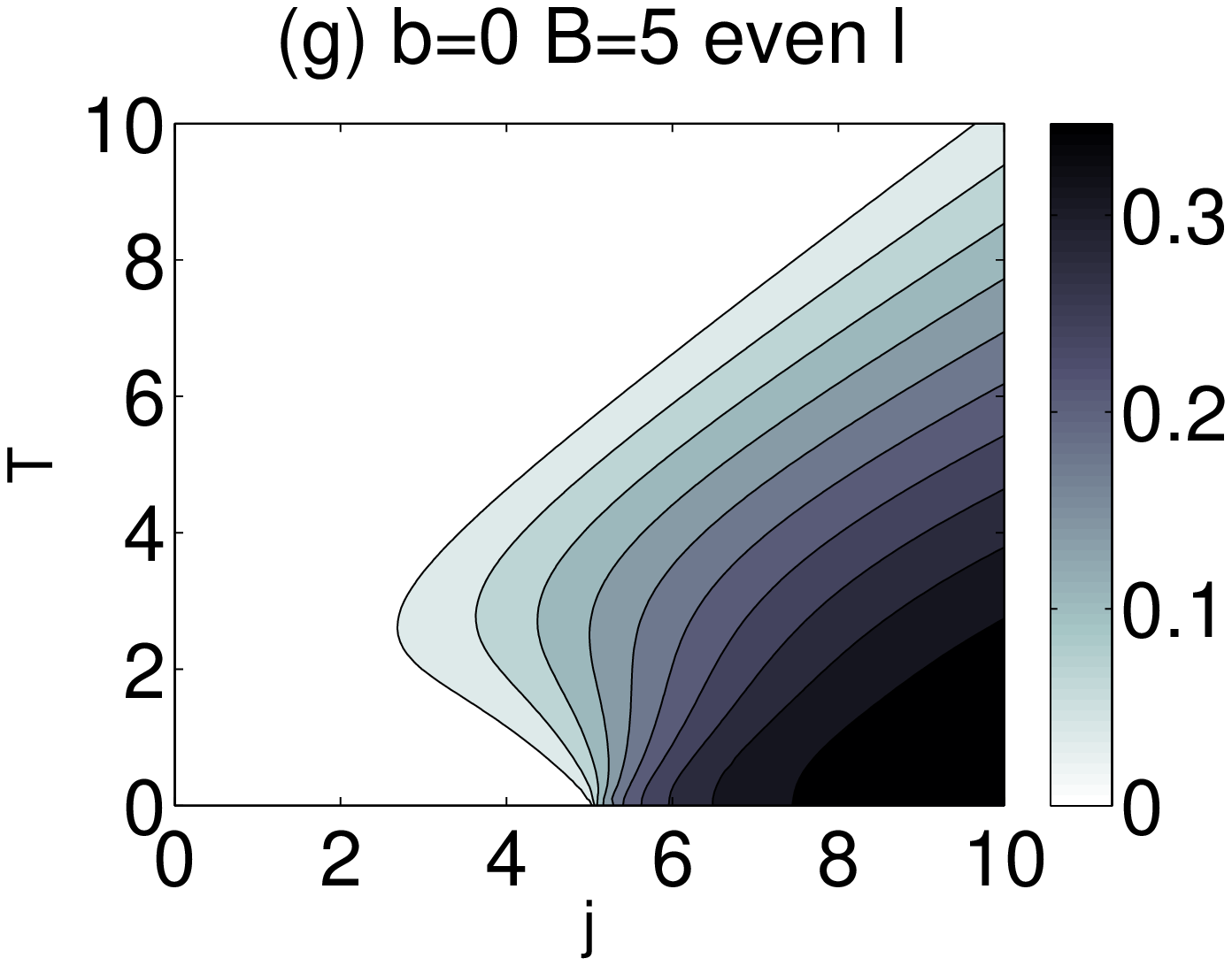,width=0.5\columnwidth,clip=} &
\epsfig{file=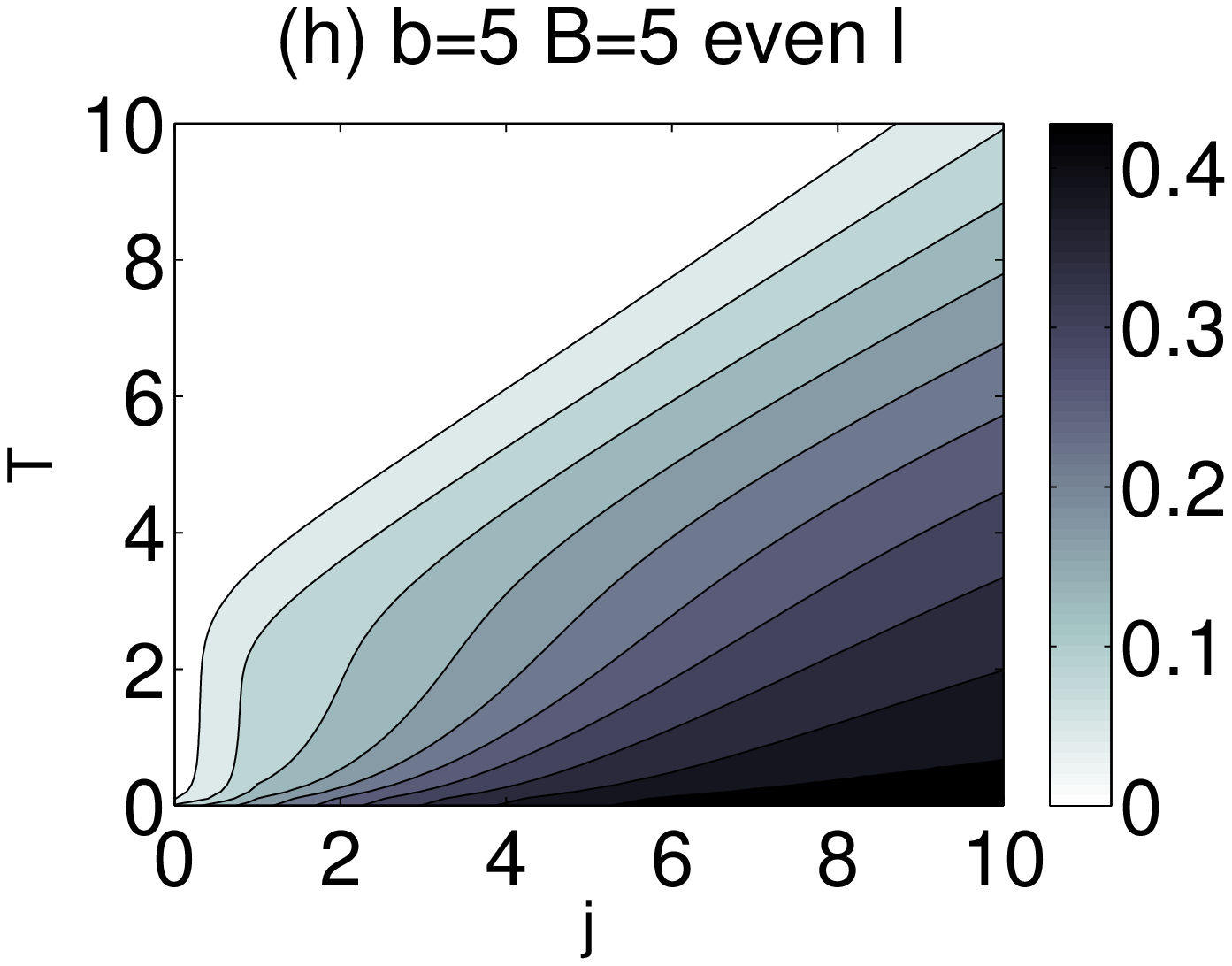,width=0.5\columnwidth,clip=}
\end{tabular}
\caption{Nearest neighbour concurrence for odd and
even sites.
 }
\label{Fig:NNConcT1}
\end{figure}

We next investigate the entanglement properties of the thermal states of the Hamiltonian
using the concurrence.

\begin{figure}[t]
\centering
\begin{tabular}{cc}
\epsfig{file=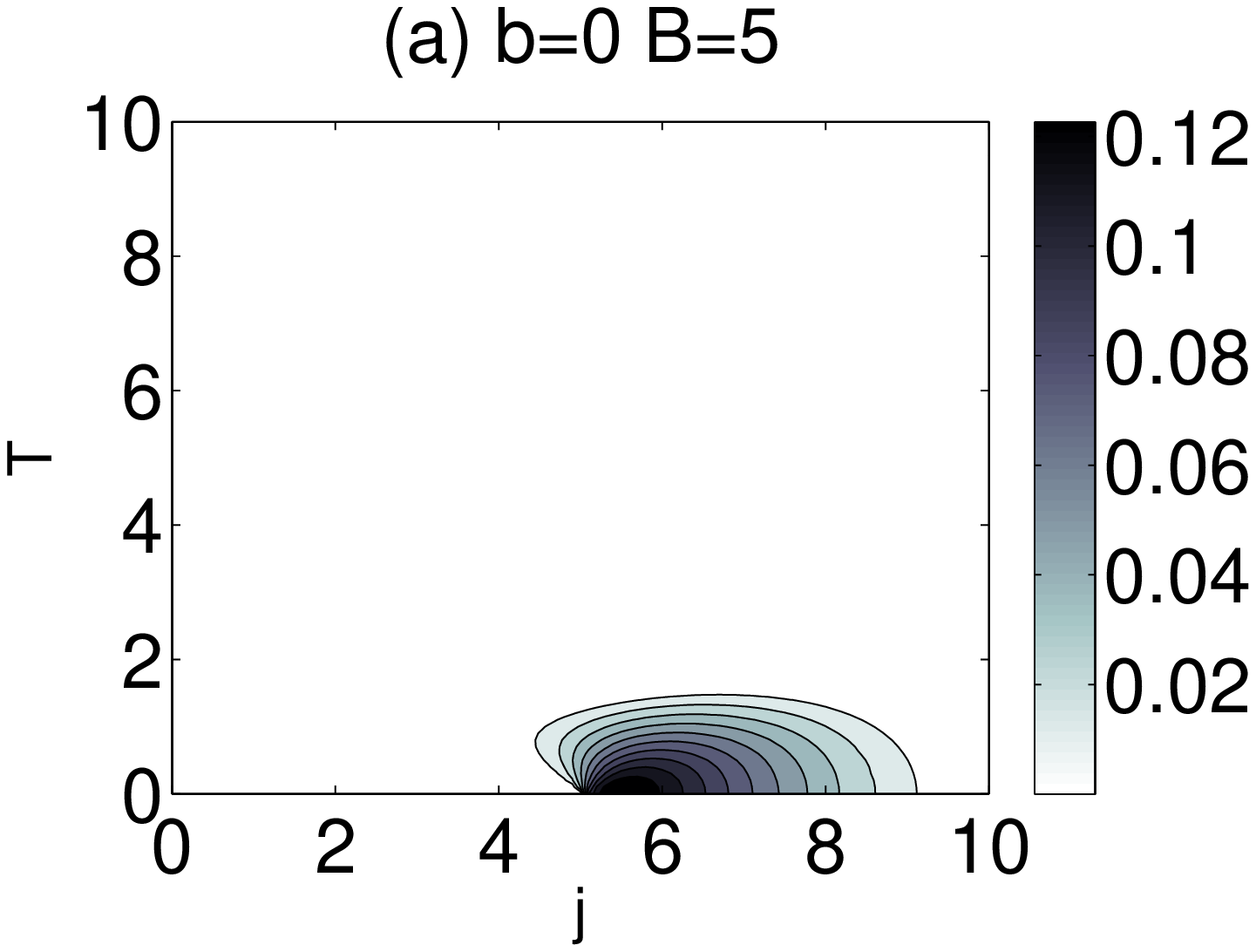,width=0.5\columnwidth,clip=} &
\epsfig{file=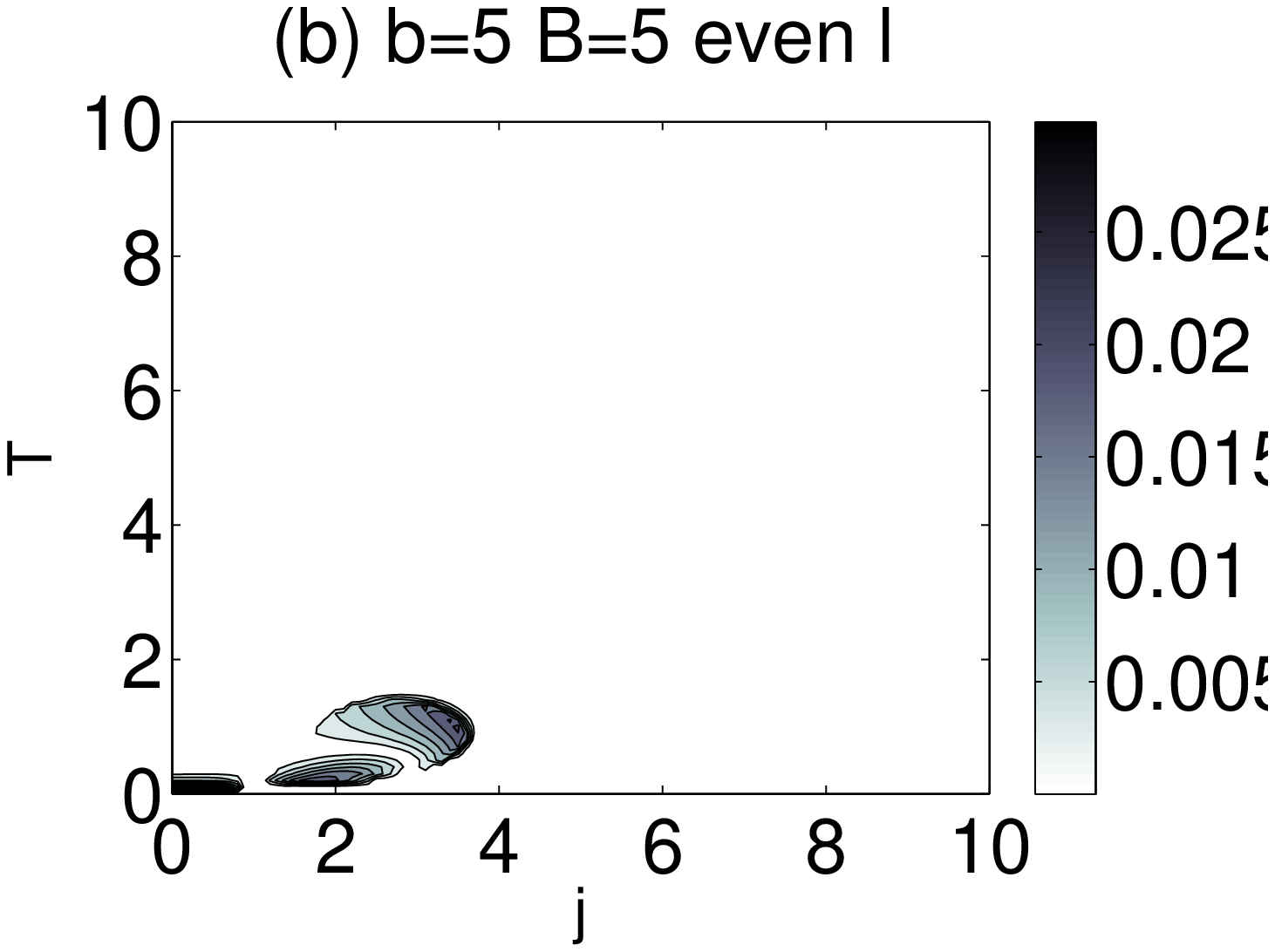,width=0.5\columnwidth,clip=}
\end{tabular}
\caption{Next nearest neighbour concurrence. When $b=0$
and $B=5$, (a), the plot for odd and even sites is identical.
No next nearest neighbour concurrence
is found for the other values of $B$ and $b$ shown for nearest neighbour
concurrence.}
\label{Fig:NNNConcT1}
\end{figure}

Fig.~\ref{Fig:NNConcT1} plots the nearest neighbour concurrence for both
odd and even sites for varying temperature and alternating coupling strength, $j$.
The figures show that increasing $j$ allows the spin chain to be entangled
at higher temperatures, and that increasing both $b$ and $B$ can enlarge the
region of entanglement. That is, the spin chain is entangled for more values of
$T$ and $j$ for higher $b$ and $B$. This is true for even as well as odd sites.
Next nearest neighbour concurrence can be seen in Fig.~\ref{Fig:NNNConcT1} where
we again see that increasing the magnetic fields can be beneficial to entanglement.
As discussed previously, there is no entanglement at $j=1$ for odd sites at any
temperature.

Increasing temperature has the effect of mixing energy
levels, something which has the ability to either
increase or decrease entanglement, though a high enough
temperature will destroy entanglement. Increasing the alternating coupling
strength counteracts this to some extent, though a high enough temperature
will still destroy the concurrence.

We note that increasing $b$ allows larger values of concurrence at higher
temperatures at lower, more accessible values of $j$ as demonstrated in
Fig.~\ref{Fig:NNConcT1} (b) and (f). Thus it is the combination
of alternating fields $b$ and $j$ that allow the spin chain to be entangled
at higher temperatures. Increasing $B$ has a similar effect, though to a
lesser extent.

%%%%%%%%%%%%%%%%%%%%%%%%%%%%%%%%%%%%%%%%%%%%%%%%%%%%%%%%%%%%%%%%%%%%%%%%%%%%%%%%%%%%%%%
\subsection{An entanglement witness}
%%%%%%%%%%%%%%%%%%%%%%%%%%%%%%%%%%%%%%%%%%%%%%%%%%%%%%%%%%%%%%%%%%%%%%%%%%%%%%%%%%%%%%%

In order to detect rather than measure entanglement in this system, we use an entanglement
witness based on the expectation value of the Hamiltonian:

\begin{equation}
\frac{4 | U + BM + bM_s |}{N(|J-j|+|J+j|)} \leq 1,
\label{Eq:EnTWit1}
\end{equation}
where $U$ is the internal energy (Eq.~\ref{Eq:InternalEn}) $M$ is the
magnetisation (Eq.~\ref{Eq:Magnetisation1}) and $M_s$ is the staggered
magnetisation (Eq.~\ref{Eq:StaggeredMag}).
The bound is found similarly to the usual method \cite{ent_wit,vlat}. Rearranging
the expectation value of the Hamiltonian gives $2|U+BM+bM_s| = |J \sum_l \langle
\sigma_l^x \sigma_{l+1}^x + \sigma_l^y \sigma_{l+1}^y \rangle +j \sum_l e^{i\pi l} \langle
\sigma_l^x \sigma_{l+1}^x + \sigma_l^y \sigma_{l+1}^y \rangle |$. The absolute sign
allows us to write $4|U+BM+bM_s|/N \leq |(J-j) | |\langle
\sigma_l^x \sigma_{l+1}^x + \sigma_l^y \sigma_{l+1}^y \rangle|_{l,odd} +|J+j| |\langle
\sigma_l^x \sigma_{l+1}^x + \sigma_l^y \sigma_{l+1}^y \rangle |_{l,even}$. Next, the
bound for both the odd and even $l$ for pure product states can be found
using the Cauchy-Schwarz inequality, and the definition of the density matrix
giving
$|\langle \sigma_l^x \sigma_{l+1}^x + \sigma_l^y \sigma_{l+1}^y \rangle| \leq 1$.
Due to the convexity of the set of separable states, this bound is also true
for all separable states while an entangled state can violate this bound.

 \begin{figure}[t]
 \begin{center}
 \centerline{
 \includegraphics[width=3.5in]{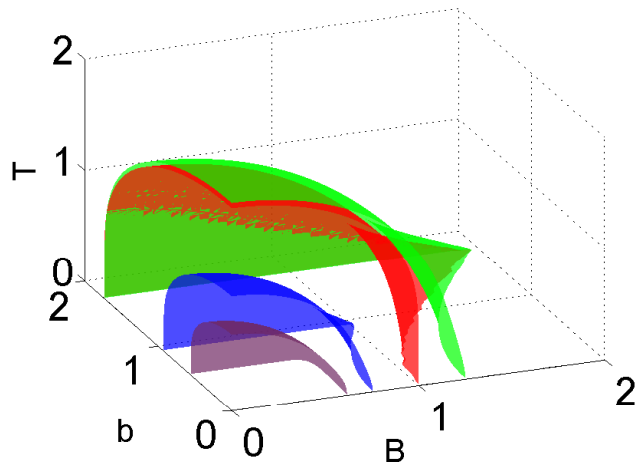} }
 \end{center}
 \caption{We witness the entanglement at different values of the alternating
 coupling strength, $j$ for $j=0$ purple, $j=0.5$ blue, $j=1$ red, $j=1.5$ green,
 from inside to out.}
 \label{Fig:Witness1}
 \end{figure}

Fig.~\ref{Fig:Witness1} demonstrates that increasing the alternating coupling
strength increases the region of entanglement detected by the witness. That is, at
larger $j$, entanglement is detected for higher values of $B$, $b$ and $T$ than is
possible at smaller $j$. This trend persists even at very high values of $j$.
In addition, the alternating magnetic field increases the maximum values of $B$
for which entanglement is detected. However, overall, increasing $B$, $b$ or $T$
enough (except at $B \sim b$ as discussed in Section~\ref{SubSec:Meyer})
will destroy entanglement either by causing the spins to align with the
magnetic field, or via the mixing of energy levels as the temperature is raised.

The entanglement witness generally complements the results of the entanglement measures, and
allows for the possibility of detecting multipartite entanglement that cannot be
measured by them. However, for our Hamiltonian, comparing Fig.~\ref{Fig:Witness1}
to Figs.~\ref{Fig:NNConcT1} and~\ref{Fig:NNNConcT1}, it can be seen that
this witness does not detect any extra entangled regions compared to the
concurrence.

%%%%%%%%%%%%%%%%%%%%%%%%%%%%%%%%%%%%%%%%%%%%%%%%%%%%%%%%%%%%%%%%%%%%%%%%%%%%%%%%%%%%
\section{Concluding Remarks}
%%%%%%%%%%%%%%%%%%%%%%%%%%%%%%%%%%%%%%%%%%%%%%%%%%%%%%%%%%%%%%%%%%%%%%%%%%%%%%%%%%%%
\label{Sec:Con}

We have found that the introduction of an alternating coupling strength and alternating
magnetic field into the usual $XX$ spin chain in a uniform magnetic field can,
for certain values of the parameters,
increase both the amount and region of entanglement quantified by either the Meyer-Wallach
measure or the concurrence. This is the case for both zero and finite temperatures.
We have demonstrated that two quantum phase transitions exist
in this system, signs of which are evident in both entanglement measures. In addition, we
have calculated an entanglement witness which detects entanglement within a region
which agrees with the measures of entanglement we consider.

It would be interesting to calculate the finite temperature effects of the
quantum phase transitions in this model.
Determining the effect on entanglement of increasing the
period of the staggered parameters would also be an interesting extension to this work. 
For example, by varying the magnetic field and coupling strength over
three sites $l$, $l+1$ and $l+2$ rather than the two considered here. However,
this may not be possible to do analytically.

\begin{acknowledgments}
This work was supported by Project for Developing Innovation Systems of
the Ministry of Education, Culture, Sports, Science and Technology (MEXT),
Japan. J.~H. acknowledges support from the JSPS postdoctoral fellowship 
for North American and European Researchers (short term).
Y.~N. acknowledges support from JSPS by KAKENHI (Grant No. 222812)
and M.~M. acknowledges support from JSPS by KAKENHI  (Grant No. 23540463).
\end{acknowledgments}

\appendix

\section{Calculations of quantities at zero temperature}
\label{App:ZeroTemp}

Here, we give exact expressions of the ground energy, the magnetization at zero temperature
and the Meyer-Wallach measure of the ground state.
For simplicity, we define regions such as
\begin{align}
\mathcal{B}_1 &:= [0, \sqrt{J^2+b^2}), \\
\mathcal{B}_2 &:= [\sqrt{J^2+b^2}, \sqrt{j^2+b^2}), \\
\mathcal{B}_3 &:=[\sqrt{j^2+b^2}, \infty), \\
\mathcal{B}_3' &:=[\sqrt{J^2+b^2}, \infty)
\end{align}

\subsection{Ground energy}

First, we show the exact expressions of the ground energy $\epsilon_g$ given by Eq.~\eqref{Eq:GroundEnergy}.

\subsubsection*{\underline{$j<J$}}
In this case, $Q$ is given by
\begin{equation}
Q = (0, \Xi) \cup   (\pi - \Xi, \pi).
\end{equation}
It is straightforward to calculate the ground energy;
\begin{equation}
\epsilon_g =
\begin{cases}
(\frac{2}{\pi} \Xi -1) B - \frac{2}{\pi} \int_0^{\Xi} \Theta(q) dq, &  \text{for } B \in \mathcal{B}_1 \\
- B, &  \text{for } B \in \mathcal{B}_3'.
\end{cases}
\end{equation}

\subsubsection*{\underline{$j=J$}}
Since $\Lambda_q^-=2B - 2 \sqrt{J^2 + b^2}$, the ground energy is obtained as
\begin{equation}
\epsilon_g =
\begin{cases}
-\sqrt{J^2 + b^2}, & \text{for } B \in \mathcal{B}_1 \\
- B, & \text{for } B \in \mathcal{B}_3'.
\end{cases}
\end{equation}

\subsubsection*{\underline{$j>J$}}
In this case, $Q$ is given by
\begin{equation}
Q = (\Xi, \pi - \Xi).
\end{equation}
Then, the ground energy is calculated as
\begin{equation}
\epsilon_g =
\begin{cases}
- \frac{2}{\pi} \int_0^{\pi/2} \Theta(q) dq, & \text{for } B \in \mathcal{B}_1\\
-\frac{2}{\pi} \Xi B - \frac{2}{\pi} \int_{\Xi}^{\pi/2} \Theta(q) dq, & \text{for } B \in \mathcal{B}_2\\
- B, & \text{for } B \in \mathcal{B}_3.
\end{cases}
\end{equation}

\subsection{Magnetization at zero-temperature}

We show the magnetization at zero temperature per site $m_g$.
The magnetization $m_g$ is directly obtained from Eq.~\eqref{Eq:Magnetisation1} such as
\begin{align}
m_g &=\lim_{\beta \rightarrow \infty} \int_0^{\pi} \frac{dq}{2 \pi} [\tanh(\beta \Lambda^+) + \tanh(\beta \Lambda^-)]\\
&= \int_{q \notin Q} \frac{dq}{\pi}.
\end{align}
By substituting $Q$, the magnetization is obtained as follows;
for $j<J$
\begin{equation}
m_g =
\begin{cases}
1- \frac{2}{\pi} \Xi, & \text{for } B \in \mathcal{B}_1\\
1, & \text{for } B \in \mathcal{B}_3',
\end{cases}
\end{equation}
for $j=J$,
\begin{equation}
m_g =
\begin{cases}
0, & \text{for } B \in \mathcal{B}_1 \\
1, & \text{for } B \in \mathcal{B}_3',
\end{cases}
\end{equation}
and, for $j>J$,
\begin{equation}
m_g =
\begin{cases}
0, & \text{for } B \in \mathcal{B}_1\\
\frac{2}{\pi} \Xi, & \text{for } B \in \mathcal{B}_2\\
1, & \text{for } B \in \mathcal{B}_3.
\end{cases}
\end{equation}

\subsection{Meyer-Wallach measure of the ground state}

Here, we give the exact expressions of the Meyer-Wallach measure of the ground state, $E_{\rm MW}$, given by Eq.~\eqref{Eq:EMWGround}.
For $j<J$,
\begin{equation}
E_{\rm MW} =
\begin{cases}
\frac{4}{\pi} \Xi (1- \frac{\Xi}{\pi}) - ( \frac{2b}{\pi} \int_0^{\Xi} \frac{1}{\Theta(q)}dq )^2 & \text{for } B \in \mathcal{B}_1\\
0, & \text{for } B \in \mathcal{B}_3'.
\end{cases}
\end{equation}
For $j=J$,
\begin{equation}
E_{\rm MW} =
\begin{cases}
\frac{J^2}{J^2+ b^2} & \text{for } B \in \mathcal{B}_1\\
0 & \text{for } B \in \mathcal{B}_3'.
\end{cases}
\end{equation}
Finally, for $j>J$,
\begin{equation}
E_{\rm MW} =
\begin{cases}
1-(\frac{2b}{\pi} \int_0^{\pi/2} \frac{1}{\Theta(q)} dq)^2 & \text{for } B \in \mathcal{B}_1\\
1- \frac{4\Xi^2}{\pi^2} - (\frac{2b}{\pi} \int_{\Xi}^{\pi/2} \frac{1}{\Theta(q)} dq)^2 & \text{for } B \in \mathcal{B}_2\\
0 & \text{for } B \in \mathcal{B}_3.
\end{cases}
\end{equation}

\end{document}